\documentclass[aps,pra,reprint,superscriptaddress,letterpaper,twocolumn,floatfix]{revtex4-2}
\usepackage{amsmath,amssymb,mathtools,dsfont,esint,textgreek,soul}

\usepackage{graphicx,setspace}
\usepackage[dvipsnames, x11names]{xcolor}
\usepackage[colorlinks]{hyperref}

\newcommand{\cqa}{_{\rm cqa}}

\hypersetup{
 colorlinks=true,
 citecolor=red,
 linkcolor=blue,
 urlcolor=cyan}

\begin{document}

\title{Exact steady states of interacting driven dissipative fermionic systems with hidden time-reversal symmetry}

\date{\today}
\author{Andrew Lingenfelter}
\email{andrew.lingenfelter@uibk.ac.at}
\affiliation{Pritzker School of Molecular Engineering, University of Chicago, Chicago, IL 60637, USA}
\affiliation{Department of Physics, University of Chicago, Chicago, IL 60637, USA}
\author{Aashish A. Clerk}
\affiliation{Pritzker School of Molecular Engineering, University of Chicago, Chicago, IL 60637, USA}

\begin{abstract}
We present exact solutions for the non-equilibrium steady states of a class of dissipative spinless fermionic systems with arbitrary Hamiltonian pairing terms, global charging energy interactions, and uniform single particle loss on every site. 
Our exact solution is found by generalizing the coherent quantum absorber technique to fermionic systems, and our result establishes the existence of hidden time-reversal symmetry in driven-dissipative fermionic models. 
The steady state exhibits a first order phase transition in the particle density, with the resulting jump discontinuity in density persisting even for finite dissipation rates.
A mean-field description of the model exhibits a bistable regime that encompasses the first-order transition line yet which fails to accurately predict its precise location via a Maxwell construction. 
We also show that the model's hidden time-reversal symmetry results in an Onsager symmetry of certain two-time correlation functions.
\end{abstract}

\maketitle


\section{Introduction}


Many-body systems subject to both coherent dynamics and dissipation often exhibit non-equilibrium dynamics and are typically driven toward non-equilibrium steady states.
The recent experimental progress in trapping and manipulating cold fermionic gases as well as the trapping of fermionic atoms in lattices has spurred theoretical study of non-equilibrium dissipative fermionic systems including, for example, the stabilization of non-equilibrium states such as BECs of paired fermions \cite{diehl_Quantum_2008} or states with nontrivial topology \cite{diehl_Topology_2011,hamanaka_Interactioninduced_2023}.
The dynamics of dissipative fermionic systems have been investigated in a variety of contexts \cite{buchhold_Nonequilibrium_2015,mazza_Dissipative_2023,soares_Dissipative_2025,shibata_Dissipative_2019,vanhoecke_Dissipative_2025,starchl_Relaxation_2022,starchl_Quantum_2024,picano_Heating_2025,borchia_InteractionMediated_2026}.

The non-equilibrium  steady states of dissipative many-body systems can offer insight into the competition between coherent dynamics and dissipation, and exact solutions of interacting systems are particularly useful for this task.
For example, Ref.~\cite{borchia_InteractionMediated_2026} elucidated intriguing dynamics caused by the interplay between dissipation-induced nonreciprocity and interactions via a fermion model that is exactly solvable due to the block-diagonal nature of its Lindbladian in momentum space \cite{buca_note_2012,mcdonald_Exact_2022}.
A recent technique for finding exact non-equilibrium steady states, the coherent quantum absorber (CQA) method \cite{roberts_Hidden_2021,stannigel_Drivendissipative_2012}, has proved extremely useful for finding the steady states of several interacting driven-dissipative bosonic and spin models.
These have included a single driven Kerr oscillator \cite{roberts_DrivenDissipative_2020}, many-body bosonic lattices \cite{roberts_Competition_2023}, a dissipative transverse-field Ising model \cite{roberts_Exact_2023}, and interacting boundary-driven spin chains \cite{lingenfelter_Exact_2024,yao_Hidden_2025}.
The solvability of these models via CQA is directly implied by the existence of a hidden time-reversal symmetry (hTRS) \cite{roberts_Hidden_2021}.
This notion of time-reversal symmetry generalizes conventional notions of TRS in open systems to a broader class of models, and it implies an Onsager symmetry of a class of two-time steady state correlation functions \cite{roberts_Hidden_2021}.
More recently, the existence of hTRS in a model has been used to estimate decoherence and switching rates in a Kerr oscillator \cite{carde_Nonperturbative_2026,mylnikov_Qubit_2025}, thus, illustrating the utility of hTRS beyond steady state properties of the model.
Given the success of CQA in finding steady states of spin and bosonic systems, it is natural to ask whether it can be applied to dissipative fermionic models, and thus, whether fermionic systems can have hTRS.

\begin{figure}[t!]
	\centering
	\includegraphics[]{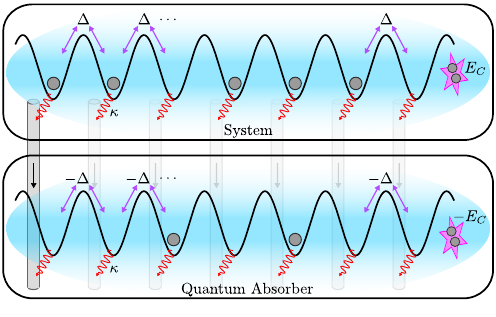}
	\caption{
	\textbf{Schematic of the 1D p-wave pairing model and CQA construction of the doubled system.}
    The system (top) consists of a 1D lattice of spinless fermions at a chemical potential $\mu$ subject to nearest neighbor pairing $\Delta$ and an all-to-all charging energy interaction $E_C$. Each lattice site has single particle loss with rate $\kappa$.
    The quantum absorber (bottom) is a mirrored copy of the system with chemical potential $-\mu$, nearest-neighbor pairing amplitude $-\Delta$ and charging energy $-E_C$.
    The system and absorber are coupled through the dissipation on each lattice site mediated by a unidirectional waveguide.
    }
	\label{fig:intro}
 \end{figure}

Two potential complications arise when considering the hTRS of fermionic systems. 
First, Kramers' theorem, as applied to open systems, allows for the the possibility of degenerate steady states and a time-reversal operator satisfying $\hat T^2 = -1$ \cite{lieu_Kramers_2022}.
However, the known hTRS systems with multiple steady states have distinct TRS operators $\hat T_i$ for distinct steady states, with each operator satisfying $\hat T_i^2  = 1$ \cite{roberts_Hidden_2021}.
We leave the resolution of this tension to future works.
Second, the CQA construction that defines hTRS requires the introduction of an auxiliary absorber system.
This poses no problems for spins or bosons, as their composite systems are composed under a simple tensor product of the Hilbert spaces.
Composite fermionic systems, however, are composed under the exterior product, for which the notion of truly isolated subsystems does not exist; fermions of different isolated subsystems must still anticommute.
This leads to the natural question: can the steady states of any fermionic systems be found using a suitable generalization of the CQA construction?

In this paper we answer this question in the affirmative, and thereby extend the notion of hidden time-reversal symmetry to open fermionic systems. 
We show that under fairly general conditions, the CQA technique can be extended to interacting fermionic models, and we present a class of such models: interacting spinless fermions with single particle loss on every site and coherent pairing between arbitrary pairs of sites.
We specialize our discussion to a dissipative Kitaev wire with nearest neighbor $p$-wave pairing \cite{kitaev_Unpaired_2001}, uniform single fermion loss on all sites, and a global charging energy.
The interplay between pairing and global charging energy has been studied in closed system model \cite{lapa_Rigorous_2020}, and the interplay between interactions and dissipation has been studied in the absence of Hamiltonian pairing \cite{buchhold_Nonequilibrium_2015,mazza_Dissipative_2023,soares_Dissipative_2025}.
Combining interactions, dissipation, and coherent driving gives us a minimal model with all of the necessary ingredients to obtain a nontrivial non-equilibrium steady state which depends simultaneously on all three ingredients.

We find an exact analytic expression for the non-equilibrium steady state of this model, and we prove that in the thermodynamic limit the exact steady state has a first order phase transition between a low-density and a high density phase.
Moreover, the mean-field description of this model exhibits a bistable regime surrounding the first order transition line in the phase diagram.
Remarkably, we find that this mean-field theory can be mapped to the mean-field theory of a dissipative transverse-field Ising model with dephasing and an inhomogeneous transverse field, via the Anderson pseudospin mapping \cite{anderson_RandomPhase_1958}.
Despite there being no direct mapping of the fermion single particle loss to the spin model, we find that a suitable combination of spin loss and dephasing yields an exact mapping of the dynamics of expectation values between the spin operators and their fermionic equivalents.
Although this mapping is not exact in the fully interacting model, it could provide new approximation methods to spin models by mapping them to exactly solvable fermion models.
Finally, we demonstrate the hidden time-reversal symmetry of the fermionic model by observing the Onsager time-symmetry of certain correlation functions.


\section{Model}


We consider a system of $L$ spinless fermions with annihilation operators $\hat c_j$ ($j=1,\dots,L$) subject to a chemical potential $\mu$ and a global charging energy $E_C$, i.e., an all-to-all density-density interaction.
We suppose there is some bulk superconductor in close proximity that induces pairing within the system, but we let the form of the pairing be completely general.
Thus, the general Hamiltonian is
\begin{align}
    \hat H = &-\mu \sum_{j=1}^{L} \hat n_j + \frac{E_c}{2L}\Big(\sum_j \hat n_j\Big)^2
    + \sum_{ij} (\Delta_{ij}\hat c_i^\dagger \hat c_{j}^\dagger + {\rm H.c.}),
    \label{eq:H-general}
\end{align}
where $\Delta_{ij} = -\Delta_{ji}$ is the antisymmetric pairing matrix, which
in general, breaks both translation invariance and particle number conservation.
Notice that the model has no notion of dimensionality; the form of $\Delta_{ij}$ can be chosen to define a $d$-dimensional lattice, but as we will show, an exact solution exists for an arbitrary $\Delta_{ij}$ matrix.
Here, we have not included any tunneling as we want to study a model with the minimal number ingredients (i.e., coherent driving, nonlinearity, and dissipation) that is solvable and which has a nontrivial non-equilibrium steady state.
Moreover, the inclusion of tunneling breaks the solvability of this model (and hence its hTRS) when $E_C \neq 0$.

The dissipation is uniform single particle loss on every site with rate $\kappa$, which gives rise to the fermionic Lindbladian 
\begin{align}
    \partial_t \hat \rho = \mathcal{L} \hat\rho \equiv -i[\hat H,\hat\rho] + \kappa\sum_{j=1}^{L} \mathcal{D}[\hat c_j]\hat \rho, \label{eq:qme}
\end{align}
where the Lindblad dissipator is $\mathcal{D}[\hat O]\hat\rho = \hat O \hat\rho \hat O^\dagger - \{\hat O^\dagger \hat O,\hat\rho\}$.
The dynamics of this model are governed by the competition of the pairing terms $\propto\Delta_{ij}$ that coherently inject correlated pairs of fermions into the system with the charging energy interaction $\propto E_C$, which creates an energy cost for each additional particle pair injected, and the dissipation $\propto \kappa$, which incoherently removes particles and destroys the pair correlations.
In a cold atom lattice, the particle loss dissipation could be realized simply as atom loss on each site in the lattice.
In a superconducting wire, it could be realized by weakly coupling the wire to a nearby fermionic reservoir (e.g., a piece of bulk normal metal) with a very low chemical potential $\mu_{\rm bath}\to -\infty$ such that electrons in the wire tunnel into unoccupied modes of the bath.

We are interested in the long-time non-equilibrium steady state of this model, $\mathcal{L}\hat\rho_{\rm ss} = 0$, which we assume to be unique due to the single particle loss on every site. 
In spin and bosonic systems, the uniqueness of the steady state is guaranteed if there is loss on every site \cite{nigro_uniqueness_2019}. 
Although the theorem of Ref.~\cite{nigro_uniqueness_2019} does not formally apply to the fermionic model, numerical exact diagonalization of small systems finds no evidence for multiple steady states in any parameter regime.

As we show in App.~\ref{app:cqa-general-soln}, for any antisymmetric pairing matrix $\Delta_{ij}$, and for all values of the parameters $\mu$, $E_c$, and $\kappa>0$, we can find the exact long-time steady state $\rho_{\rm ss}$.
To be concrete, in the remainder of the paper we focus on the specific case of uniform nearest neighbor $p$-wave pairing in 1D (see Fig.~\ref{fig:intro}): $\Delta_{ij} = \Delta(\delta_{i,j-1} - \delta_{i,j+1})/2$, such that the Hamiltonian becomes
\begin{align}
    \hat H = &-\mu \sum_{j=1}^{L} \hat n_j + \frac{E_c}{2L}\Big(\sum_j \hat n_j\Big)^2 \label{eq:H-1D}
    + \Delta\sum_{j} (\hat c_j^\dagger \hat c_{j+1}^\dagger + {\rm H.c.}).
\end{align}
We assume periodic boundary conditions (PBC), $\hat c_{L+1}\equiv \hat c_{1}$, and for further simplicity we always choose an even number of sites, $L=2m$.


\section{Coherent quantum absorber exact solution}


The basic idea of the coherent quantum absorber (CQA) solution method is to introduce a fictitious second ``absorber'' system $B$ with a Hilbert space isomorphic to that of the original system $A$ and with the same number of Lindblad dissipators.
We construct a cascaded composite system by modeling each dissipator $\hat L_{j,A}$ of system $A$ as directional (chiral) waveguide. Then, each dissipator $\hat L_{j,B}$ of system $B$ is coupled to the $j^{\rm th}$ waveguide downstream of system $A$, so that excitations which leak out of system $A$ interact with the absorber $B$ (see Fig.~\ref{fig:intro}) \cite{gardiner_Driving_1993,carmichael_Quantum_1993}.
If the Hamiltonian and jump operators of the absorber are chosen correctly, then the composite CQA system relaxes to a \emph{pure} entangled state $|\psi\cqa\rangle$ of the system $A$ and absorber $B$ \cite{roberts_Hidden_2021,stannigel_Drivendissipative_2012}.
Because the CQA system is cascaded, the dynamics of the absorber have no influence on the dynamics of system $A$; hence, the steady state of system $A$ is obtained by tracing out the absorber $\hat\rho_{\rm ss} = {\rm Tr}_B(|\psi\cqa \rangle\langle \psi\cqa|)$.
For a system with hidden time-reversal symmetry (hTRS), the absorber system is a ``perfect absorber'' whose Hamiltonian and dissipation mirror that of the original system: $\hat H_B = -\hat H_A$ and $\hat L_{j,B} = -\hat L_{j,A}$  \cite{roberts_Hidden_2021}. 
We find that Eq.~\eqref{eq:qme} has a perfect absorber and, therefore, is the first example of a fermionic system with hTRS.
However, as we noted in the introduction, the anticommuting nature of fermions requires that composite fermionic systems be constructed using the exterior product, and there are subtleties to the creation of a cascaded fermionic system which we address next.

\subsection{Cascaded fermionic dissipation with odd jump operators}

For bosonic and spin systems with arbitrary cascaded interactions, it is straightforward to construct the cascaded CQA system following the standard recipe \cite{metelmann_Nonreciprocal_2017}.
The standard cascade construction assumes the system $A$ dissipators $\hat L_{j,A}$ commute with all system $B$ operators, $[\hat L_{j,A},\hat O_B]=0$, and vice versa \cite{carmichael_Quantum_1993,gardiner_Driving_1993,metelmann_Nonreciprocal_2017}.
The cascaded CQA master equation is
\begin{align}
	\partial_t \hat\rho\cqa &= -i[\hat H\cqa,\hat\rho\cqa]  + \sum_j \mathcal{D}[\hat L_{j,{\rm cqa}}]\hat\rho\cqa,  \label{eq:cqa-qme} 
\end{align}
where $\hat H\cqa = \hat H_A - \hat H_B - (i/2)\sum_j (\hat L_{j,A}^\dagger\hat L_{j,B} - {\rm H.c.})$ and $\hat L_{j,{\rm cqa}} = \hat L_{j,A} - \hat L_{j,B}$. 
Here the $\propto -i/2$ term is the unidirectional coupling between system and absorber.
The pure steady state $|\psi\cqa\rangle$ satisfies the dark state conditions
\begin{align}
	\hat H\cqa|\psi\cqa\rangle =\hat L_{j,{\rm cqa}}|\psi\cqa\rangle = 0,
	\label{eq:dark-cond}
\end{align}
for every collective dissipator $\hat L_{j,{\rm cqa}}$.
Note that this argument works without modification for bilinear fermionic jump operators (e.g., $\hat L_{j,A} = \hat c_{j,A}^\dagger\hat c_{j,A}$) as the conditions $[\hat L_{j,A}, \hat O_B] = [\hat L_{j,B},\hat O_A] = 0$ are satisfied \cite{malz_Current_2018}.

The situation is different for fermionic systems whose dissipators are odd in creation and annihilation operators, such as the linear dissipators in Eq.~\eqref{eq:qme}.
Here, the dissipators $\hat L_{j,s} = \sqrt{\kappa}\hat c_{j,s}$ anticommute across systems, $\{ \hat L_{i,A}, \hat L_{j,B} \} = 0$ which violates the commutator assumption.
Nevertheless, we show in App.~\ref{app:cascaded-fermionic-systems} that with minimal constraints on the Hamiltonian, a fermionic system with linear dissipators can be cascaded---\emph{in the sense that the equations of motion of all operators of system $A$ that even in creation/destruction operators are independent of system $B$}---following the standard construction.
Thus, for Eq.~\eqref{eq:qme}, the CQA Hamiltonian is of the anticipated  form $\hat H\cqa^{\rm (fermi)} = \hat H_A - \hat H_B - (i/2)\sum_j (\hat L_{j,A}^\dagger L_{j,B} - {\rm H.c.})$.

\subsection{Coherent quantum absorber}
\label{sec:cqa-setup}

We construct the coherent quantum absorber system according to Eq.~\eqref{eq:cqa-qme}.
It is convenient to work in the basis of collective ``dark'' and ``bright'' modes $\hat c_{j,\pm} = \frac{1}{\sqrt 2}(\hat c_{A,j} \pm \hat c_{B,j})$.
The dark modes $\hat c_{j,+}$ do not directly couple to dissipation, but every bright mode is damped at rate $2\kappa$, $\hat L_{j,{\rm cqa}} = \sqrt{2\kappa}\hat c_{j,-}$.

The pure steady state wavefunction $|\psi_{\rm cqa}\rangle$ must satisfy the dark state conditions Eq.~\eqref{eq:dark-cond}.
The dissipator conditions reduce to $\hat c_{j,-}|\psi\cqa\rangle=0$, thus any excitations in the wavefunction must be in the dark modes $\hat c_{j,+}$. 
Keeping only the terms of $\hat H_{\rm cqa}$ that act nontrivially on dark mode excitations, we obtain the effective non-hermitian Hamiltonian:
\begin{align}
    	\mathcal{\hat H}_{\rm cqa} &= -\sum_j \left( \tilde{\mu} - \frac{E_c}{2 L}\hat N \right)\hat c_{j,-}^\dagger \hat c_{j,+} \nonumber
	\\
	&\qquad+ \frac{\Delta}{2}\sum_{j}(\hat c_{j,+}^\dagger \hat c_{j+1,-}^\dagger - \hat c_{j+1,+}^\dagger \hat c_{j,-}^\dagger) \nonumber 
	\\
	&= \mathcal{\hat H}_0 + \mathcal{\hat H}_\Delta,
    \label{eq:H-eff}
\end{align}
where $\tilde{\mu} = \mu + i\kappa/2$ is the complex chemical potential and $\hat N = \sum_{j,s} \hat n_{j,s}$ is the total number operator. The effective Hamiltonian divides into two parts: the particle number conserving $\mathcal{\hat H}_0$, which swaps a dark mode excitation for a bright mode excitation, and the particle non-conserving $\mathcal{\hat H}_{\Delta}$ which creates a dark-bright excitation pair.

\subsection{Exact steady state solution}

We seek a wavefunction $|\psi\cqa\rangle$ comprising only dark excitations which satisfies the Hamiltonian condition $\mathcal{\hat H}_0|\psi\cqa\rangle + \mathcal{\hat H}_\Delta|\psi\cqa\rangle = 0$.
As this form of the Hamiltonian condition suggests, the steady state is a superposition determined by the term-by-term destructive interference between the action of $\mathcal{\hat H}_0$ with the action of $\mathcal{\hat H}_\Delta$ on the state (see App.~\ref{app:cqa-general-soln} for details).
For 1D $p$-wave pairing, we find that the fundamental excitation is a nearest-neighbor pair of dark mode fermions: $\hat c_{j}^\dagger\hat c_{j+1}^\dagger|0\rangle$. (Here we suppress the dark mode label $+$ for clarity.)
We define a creation operator that creates a nearest-neighbor pair which is uniformly delocalized across the chain
\begin{align}
    \hat B^\dagger = \sum_{j} \hat c_j^\dagger \hat c_{j+1}^\dagger = \frac{1}{2}\sum_k \sin k \,\hat c_k^\dagger \hat c_{-k}^\dagger.
    \label{eq:Bdagger}
\end{align}
The second expression is in momentum space (taking the lattice constant $a\equiv 1$), with $\hat c_k = L^{-1/2}\sum_j e^{ijk}\hat c_j$.
We find that the the steady state wavefunction is a power series in this pair-creation operator,
\begin{align}
    |\psi\cqa\rangle &= \sum_{n=0}^{\infty} \frac{\alpha_n}{n!} \big( \hat B^\dagger \big)^n|0\rangle, \label{eq:cqa-wavefunc}
\end{align}
where $|0\rangle$ is the fermion vacuum, $\hat c_{j,\pm}|0\rangle = 0$ and the amplitude coefficients $\alpha_n$, given below in Eq.~\eqref{eq:alpha-n-solution}, are determined by solving a recurrence relation.
Although the solution is formally an infinite series, for any finite $L<\infty$, the series terminates at $n = L/2-1$, the maximum number of pairs that can be added to an even-length PBC chain (the maximally filled state with $n=L/2$ fermions is not permitted, see App.~\ref{app:1d-p-wave}).

The destructive interference of $\mathcal{\hat H}_0$ acting on $(\hat B^\dagger)^n|0\rangle$ and $\mathcal{\hat H}_\Delta$ acting on $(\hat B^\dagger)^{n-1}|0\rangle$ defines a one-term recurrence relation for the coefficients $\alpha_n$.
The solution of this recurrence relation for $\alpha_0\equiv1$ is
\begin{align}
    \alpha_n = \frac{\Delta^n}{\prod_{m=1}^n [\tilde{\mu} - \frac{m}{L}E_c]},
    \label{eq:alpha-n-solution}
\end{align}
where $\tilde{\mu} = \mu + i\kappa/2$ the complex chemical potential.
In the free limit $E_C = 0$ they reduce to $\alpha_n = (\Delta/\tilde{\mu})^n$ and the state reduces to a Gaussian BCS state:
\begin{align}
    |\psi\cqa\rangle&_{E_C = 0} = \prod_{k>0}\left[ u_k - v_k \hat c_k^\dagger\hat c_{-k}^\dagger \right]|0\rangle,
\end{align}
where $v_k/u_k = (i\Delta/\tilde{\mu})\sin k$ and the normalization is $|u_k|^2 + |v_k|^2  = 1$.
For the interacting system, $E_C\neq 0$, each $n$-pair component of the state sees its own effective chemical potential, which modified from $\tilde{\mu}$ by the energy cost for having $n$ pairs in the system, $(n/L)E_C$.

The steady state wavefunction Eq.~\eqref{eq:cqa-wavefunc} is not normalized.
For an arbitrary pairing matrix $\Delta_{ij}$, the exact normalization is a nontrivial problem (see App.~\ref{app:cqa-general-soln}). 
Nevertheless, as we show in App.~\ref{app:1d-p-wave}, for nearest neighbor pairing in 1D, the normalization reduces to finding the number of unique combinations of $n$ nearest-neighbor hard-core dimers placed on a PBC lattice of $L$ sites: $\langle\psi\cqa|\psi\cqa\rangle = \sum_n |\alpha_n|^2 \mathcal{N}(L,n)$, where 
\begin{align}
    \mathcal{N}(L,n) = \frac{L}{n}\binom{L-n-1}{n-1}
    \label{eq:norm-pbc}
\end{align}
counts the number dimer combinations, and is defined for $n=0$ as $\mathcal{N}_{\rm PBC}(L,0)\equiv 1$.


\section{First order phase transition}
\label{sec:phase-transition}

With the exact steady state wavefunction of the CQA system, Eqs.~\eqref{eq:cqa-wavefunc} and \eqref{eq:alpha-n-solution}, in hand, we can now ask about the properties of the physical system steady state $\hat \rho_{\rm ss} = {\rm Tr}_B[|\psi\cqa\rangle\langle\psi\cqa|]$.
In the thermodynamic limit $L\to\infty$ and for chemical potential $\mu$ in the range $0 < \mu/E_C < 1/2 $, there is a first order phase transition between a low density phase for small $|\Delta| < \Delta_{\rm crit}(\mu)$ and a high-density phase for $|\Delta| > \Delta_{\rm crit}(\mu)$.
We also find that for an attractive charging energy $E_C < 0$, the phenomenology is identical but at negative chemical potential, $0 > \mu/|E_C| > -1/2$.

For what follows in the rest of this section, we fix the scale of the nonlinearity $E_C \equiv 1$ and measure $\mu$, $\Delta$, and $\kappa$ in units of $E_C$.
We also set $\Delta\geq0$, without loss of generality.

\subsection{Finite size systems}

\begin{figure*}[ht]
	\centering
	\includegraphics[width=7in]{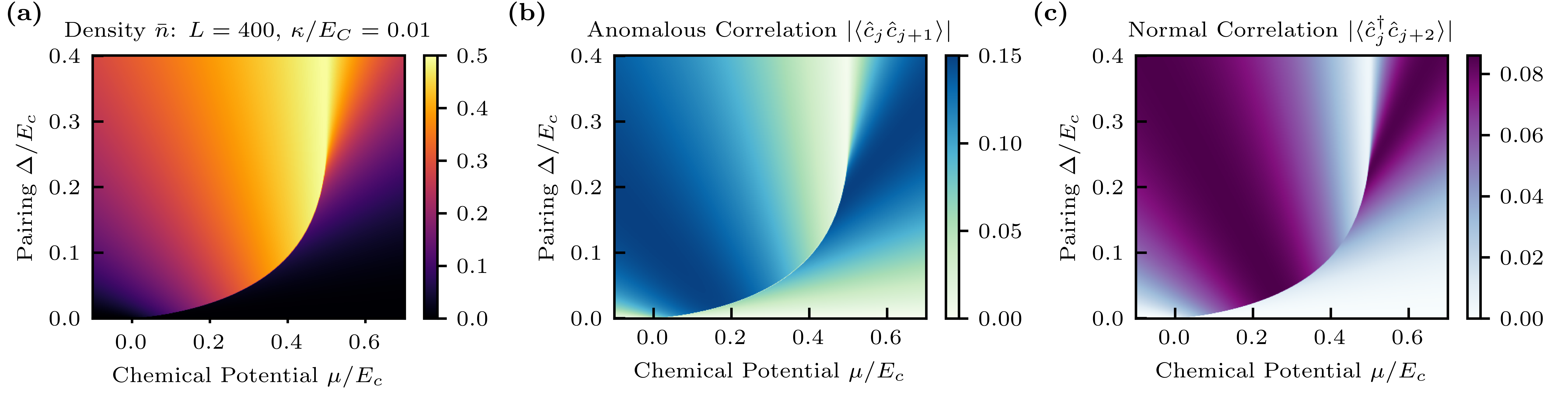}
	\caption{
	\textbf{Steady state phase diagram.}
    Phase diagram in the $\mu$--$\Delta$ plane for $L=400$ sites, charging energy $E_C \equiv 1$, and dissipation rate $\kappa/E_C = 0.01$.
    \textbf{(a)} The mean particle density $\bar{n} = L^{-1}\sum_j\langle\hat n_j\rangle$. 
    \textbf{(b)} The magnitude of the nearest neighbor anomalous correlations $|\langle\hat c_j \hat c_{j+1} \rangle|$ (translation invariance implies there is no spatial variation of the correlation in $j$). 
    \textbf{(c)} The magnitude of the nearest \emph{nonzero} normal correlations $|\langle\hat c_j^\dagger \hat c_{j+2} \rangle|$.
    \label{fig:phase-diagram}}
 \end{figure*}

From the exact solution, Eq.~\eqref{eq:cqa-wavefunc}, we numerically compute the phase diagram of particle density $\bar{n} = L^{-1}\sum_j \langle\hat n_j\rangle$ in the $\mu$--$\Delta$ plane, shown in Fig.~\ref{fig:phase-diagram}(a) for a chain with $L=400$ sites.
There is a first order phase boundary between high-density and low-density phases that terminates on one end at $\mu = \Delta =0$, and on the other end at $\mu = 0.5$ and $\Delta \approx 0.25$.
This sharp transition is also observed in the steady state anomalous correlations, e.g., $|\langle\hat c_j\hat c_{j+1}\rangle|$ (Fig.~\ref{fig:phase-diagram}(b)) and normal correlations, e.g., $|\langle\hat c_j^\dagger\hat c_{j+2}\rangle|$ (Fig.~\ref{fig:phase-diagram}(c)).

We can gain some qualitative insight into why and where the phase transition exists from the analytic form of the state vector coefficients (cf.~Eq.~\eqref{eq:alpha-n-solution})
\begin{align}
    \alpha_n = \frac{\Delta^n}{\prod_{m=1}^n [\tilde{\mu} - \frac{m}{L}]},
    \label{eq:alpha-n-EC1}
\end{align}
where we have set $E_C\equiv 1$.
Specifically, there is a ``resonance'' condition that arises from the product in the denominator. 
Recall that $\alpha_n$ is the amplitude associated with $n$ fermion pairs in the CQA wavefunction $|\psi\cqa\rangle$ (cf.~Eq.~\eqref{eq:cqa-wavefunc}).
For chemical potential $0<\mu<1/2$ (i.e., in the critical region), there is some $m_{\rm min}/L \approx \mu$ for which the product factor $|\tilde{\mu} - m_{\rm min}/L|$ is minimized among all $0\leq m \leq L/2-1$.
The coefficients $\alpha_n$ with $n\geq m_{\rm min}$ all contain this minimal ``resonant'' factor in their denominators. Thus, their magnitudes are enhanced relative to the $\alpha_{n<m_{\rm min}}$ which do not contain the resonant factor.
This results in a step-like feature in $|\alpha_n|$ at $n/L=\mu$ when plotted as as a function of $n$ \footnote{The step-like behavior of $|\alpha_n|$ is not a sharp feature for finite $\kappa$; instead, it occurs over a range of values approximately $\mu-2\kappa < n/L < \mu+2\kappa$ (in units of $E_C$).
Moreover, the functional form of the $|\alpha_n|$ can obscure the step, making it difficult to discern on a plot for generic parameters.}.
For large enough $\Delta$, this implies that the high density states with $n>m_{\rm min}$ have larger amplitudes, so the system is pushed to a high density state.
For smaller $\Delta$, the exponential suppression due to $\alpha_n \propto \Delta^n$ wins out, so the system is kept in a low density state.

The above argument suggests that for $\mu$ in the critical region, the density might change abruptly as a function of $\Delta$ around some $\Delta_{\rm crit}$, and this is borne out numerically (cf.~Fig.~\ref{fig:phase-diagram}); however, it does not guarantee a discontinuous jump in density at $\Delta_{\rm crit}$.
Next, we show that in the $L\to\infty$ limit, there is indeed a discontinuous jump in density at $\Delta_{\rm crit}(\mu)$.
Moreover, we show that the discontinuity holds for finite dissipation $\kappa>0$, not only in the weak dissipation limit $\kappa\to0$.

\subsection{Thermodynamic limit}

To take the thermodynamic limit $L\to\infty$, we first rewrite the CQA wavefunction in terms of normalized states $|n\rangle$ of $n$ fermion pairs
\begin{align}
    |\psi\cqa\rangle &= \sum_{n=0}^{L/2-1} \beta_n|n\rangle,
    \label{eq:cqa-beta}
    \\
    |n\rangle &\equiv \frac{1}{\sqrt{\mathcal{N}(L,n)}}\frac{(\hat B^\dagger)^n}{n!}|0\rangle,
    \label{eq:n-states}
    \\
    \beta_n &\equiv \alpha_n\frac{\sqrt{\mathcal{N}(L,n)}}{\sqrt{\sum_m |\alpha_m|^2 \mathcal{N}(L,m)}}
\end{align}
where $\mathcal{N}(L,n)$ is the state norm of $n$ pairs on a PBC chain of $L$ sites given by Eq.~\eqref{eq:norm-pbc}.
The states $|n\rangle$ are orthonormal, $\langle n|m\rangle = \delta_{nm}$, and the new state vector coefficients $\beta_n$ are true probability amplitudes: $\sum_n |\beta_n|^2 = 1$.
The probability to be in state $|n\rangle$ is $p_n = |\beta_n|^2$.
Note that the total particle number of a state $|n\rangle$ is $\langle n| \hat N_+ |n\rangle = 2n$, where $\hat N_+ = \sum_j \hat c_{j,+}^\dagger\hat c_{j,+}$ is total dark subspace particle number.
Therefore, $\langle n| \hat N_A |n\rangle = n$ for the physical system $A$, and the particle density of the physical system is $\bar n = \sum_n n|\beta_n|^2$. 

It is useful to define the ``density coordinate'' $\rho \equiv \frac{2n}{L}$ in the range $0\leq\rho<1$, where $\rho = 1$ is the maximally filled state.
For any finite $L$ it is assumed that $\rho$ takes the discrete values indexed by $n$, such that $|n\rangle \mapsto |\rho\rangle$ are well defined via Eq.~\eqref{eq:n-states}.
In the limit, we can treat $P(\rho) = |\beta(\rho)|^2$ as a continuous probability distribution for the particle density. Thus, the probability to measure a state with particle density within the range $\rho$ and $\rho+d\rho$ is $P(\rho)d\rho$, and the mean density is $\bar n\cqa = \int_0^1d\rho P(\rho)\rho$. Note that the \emph{physical} particle density is $\bar n = \bar n\cqa /2$.

As shown in App.~\ref{app:phase-transition}, we can define an effective free energy
\begin{align}
    Q(\rho) \equiv -\lim_{L\to\infty} L^{-1}\ln |\beta(\rho)|^2,
    \label{eq:effective-free-energy}
\end{align}
(cf.~Eqs.~\eqref{app-eq:effective-free-energy-full} and \eqref{app-eq:effective-free-energy-weak-diss} for explicit expressions), for which
the density probability distribution $P(\rho) = |\beta(\rho)|^2$ is well-approximated by $P(\rho)\propto \exp[-LQ(\rho)]$ for large but finite $L$.
In the limit $L\to\infty$, the exact probability distribution is
\begin{align}
    P(\rho)=\lim_{L\to\infty} \left[\frac{\exp \big[ -LQ(\rho) \big]}{\int_0^1d\rho \exp \big[ -LQ(\rho) \big]} \right] = \delta(\rho - \rho_{\rm min}),
    \label{eq:thermodynamic-prob-distr}
\end{align}
where $\rho_{\rm min}$ is the value that minimizes the effective free energy, $Q(\rho_{\rm min}) = \min_{\rho\in[0,1]}Q(\rho)$.
Thus, the physical particle density in the thermodynamic limit is simply $\bar n = \frac{1}{2}\int_0^1 d\rho P(\rho)\rho = \rho_{\rm min}/2$.

\begin{figure}[t!]
	\centering
	\includegraphics[]{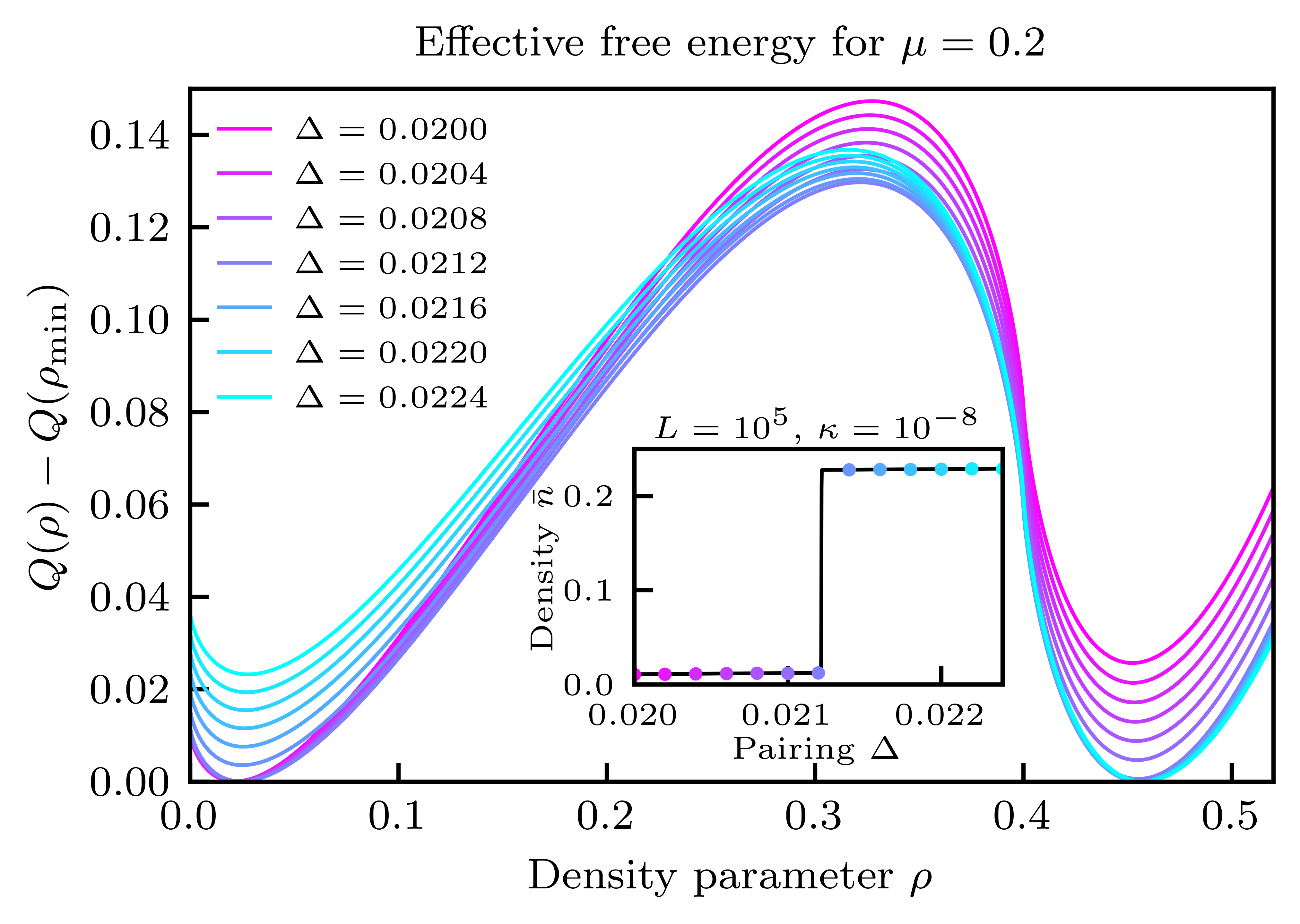}
	\caption{
	\textbf{Effective free energy and phase transition.}
    The effective free energy $Q(\rho)$ defined by Eq.~\eqref{eq:effective-free-energy} is shown at chemical potential $\mu = 0.2$ for pairing potentials ranging from $\Delta = 0.02$ to $\Delta = 0.0224$ (each successive curve steps by $0.002$).
    For each $\Delta$, the minimum $Q(\rho_{\rm min})$ is subtracted off so that each curve touches zero at exactly one point.
    The phase transition occurs at $\Delta_{\rm crit}\approx 0.02122$. 
    Note that $Q(\rho)$ for $\Delta = 0.0212$ does not quite reach zero in the high density well ($\mu\approx 0.45$).
    \textbf{Inset:}
    The exact physical particle density for a $L=10^5$ system is plotted as a function of $\Delta$ for $\mu = 0.2$ and $\kappa = 10^{-6}$ around the phase transition. Overlaid as individual points are the densities $\bar n = \rho_{\rm min}/2$ predicted from $\min Q(\rho)$ at their respective $\Delta$.
    }
	\label{fig:phase-transition}
 \end{figure}

For generic parameters $\Delta$ and $\mu$ there is a single global minimum of the effective free energy $Q(\rho)$; however, when $0<\mu<1/2$ and $\Delta>0$ is sufficiently small there are two local minima of $Q(\rho)$. 
The low density minimum is at $0<\rho_{\rm low}<2\mu$, and the high density minimum is at $2\mu<\rho_{\rm high} <1$.
For a given $\mu$ in the critical range, the phase transition occurs at pairing amplitude $\Delta_{\rm crit}(\mu)$ defined by the behavior of $Q(\rho)$:
\begin{align}
    \min Q(\rho) = 
    \begin{cases}
    Q(\rho_{\rm low}) & \Delta = \Delta_{\rm crit} - \epsilon
    \\
    Q(\rho_{\rm high}) & \Delta = \Delta_{\rm crit} + \epsilon
    \end{cases}
\end{align}
for $\epsilon>0$.
Therefore, there is a discontinuous jump in the particle density across the phase boundary in the thermodynamic limit.
This behavior is demonstrated in Fig.~\ref{fig:phase-transition}, where the effective free energy is shown in the $\kappa\to0$ limit for $\mu=0.2$. 
Below the transition, $\Delta < \Delta_{\rm crit}$, the minimum of $Q(\rho)$ is in the low density well, and above the transition, $\Delta > \Delta_{\rm crit}$, it is in the high density well.
In both cases, the particle density is $\bar n = \rho_{\rm min}/2$, which we compare to the exact solution for a large but finite system $L=10^5$ with a small but nonzero $\kappa=10^{-8}$.
The minima of $Q(\rho)$ correctly predict both the phase transition point $\Delta_{\rm crit}\approx 0.02122$ and the particle density on both sides of the transition. We stress that the value of $\rho$ at global minimum of $Q(\rho)$ \emph{always} gives the particle density $\bar n = \rho_{\rm min}/2$, not just near the transition.

Finally, we show in App.~\ref{app:phase-transition} that the discontinuous jump in particle density across the phase boundary occurs for finite dissipation.
The effect of dissipation is to cause the terminal point of the phase boundary to shift from $\mu_{\rm crit} = 1/2$ to a $\kappa$-dependent smaller value $\mu_{\rm crit} < 1/2$, and to slightly shift the phase boundary $\Delta_{\rm crit}(\mu)$ where it exists for $\mu<\mu_{\rm crit}$.
In particular, we show that for the same parameters $L=10^5$ and $\mu=0.2$ as in Fig.~\ref{fig:phase-transition}, the phase transition shifts from $\Delta_{\rm crit}\approx 0.02122$ for $\kappa\to0$ to $\Delta_{\rm crit}\approx 0.02138$ at $\kappa=10^{-3}$ (see Fig.~\ref{app-fig:phase-transition-finite-kappa}).

\section{Mean-field theory}

We compare the exact first order phase transition with the prediction from mean-field theory.
As we show in the following, the mean-field theory predicts a bistable regime of the mean particle density that encompasses the phase boundary in the $\mu$--$\Delta$ plane; however, the predicted phase boundary from the Maxwell construction is not correct.

\subsection{Mean-field treatment of charging energy}

We seek a mean-field description of the global interaction
\begin{align}
    \hat H_{\rm int} = \frac{E_C}{2L}\Big(\sum_j \hat n_j\Big)^2.
\end{align}
Unlike bosonic systems whose mean-field theories arise by assuming, e.g., $\langle \hat a_j \rangle  \neq 0$, fermionic annihilation operators do not have nonzero expectation values, $\langle \hat c_j\rangle =0$.
Instead, we assume bilinear operators (i.e., occupation numbers $\langle\hat n_j\rangle $, normal correlations $\langle\hat c_i ^\dagger\hat c_j\rangle$, and anomalous correlations $\langle \hat c_i \hat c_j \rangle$) have non-zero expectation values.
Due to translation invariance, all occupation numbers are identical and given by the average occupation $\langle\hat n_j\rangle \equiv \bar{n}$.
Thus, considering all six ``contractions'' of the interaction where one pair of operators is replaced with its mean-field expectation value, we find the mean-field interaction to be 
\begin{align}
	\hat{H}_{\mathrm{int}}^{(\mathrm{mf})}&= E_C \bar{n}\sum_{j}\hat{n}_{j}+\frac{E_C}{L}\sum_{i\neq j}\langle\hat{c}_{j}^{\dagger}\hat{c}_{i}\rangle\hat{c}_{i}^{\dagger}\hat{c}_{j} 
	\\
	&+\frac{E_C}{2L}\sum_{ij}\left[\langle\hat{c}_{j}^{\dagger}\hat{c}_{i}^{\dagger}\rangle\hat{c}_{i}\hat{c}_{j}+\mathrm{h.c.}\right]. \nonumber
\end{align}
Here we have split the normal correlations with $i\neq j$ from the density-density terms.

We compute the steady state density $\bar{n}$ by substituting the interaction for its mean-field approximation in Eq.~\eqref{eq:qme}.
As we show in App.~\ref{app:mean-field}, in the thermodynamic limit $L\to\infty$, the only relevant term in $\hat H_{\rm int}^{\rm (mf)}$ is the first term $\propto \bar{n}$:
\begin{align}
    \hat H_{\rm int}^{\rm (mf)} \to E_C\bar{n}\sum_j \hat n_j \label{eq:mf-int}
\end{align}
We obtain the self-consistency equation for the mean density
\begin{align}
	\bar{n}=\frac{1}{2}\left(1-\frac{\sqrt{(E_C\bar n-\mu)^{2}+\kappa^{2}/4}}{\sqrt{(E_C\bar n-\mu)^{2}+\kappa^{2}/4+4\Delta^{2}}}\right). \label{eq:nbar-self-cons}
\end{align}
In the noninteracting limit, $E_c = 0$, the expression is exact.

\subsection{Mean-field bistability}

\begin{figure}[t!]
	\centering
	\includegraphics[width=3.0in]{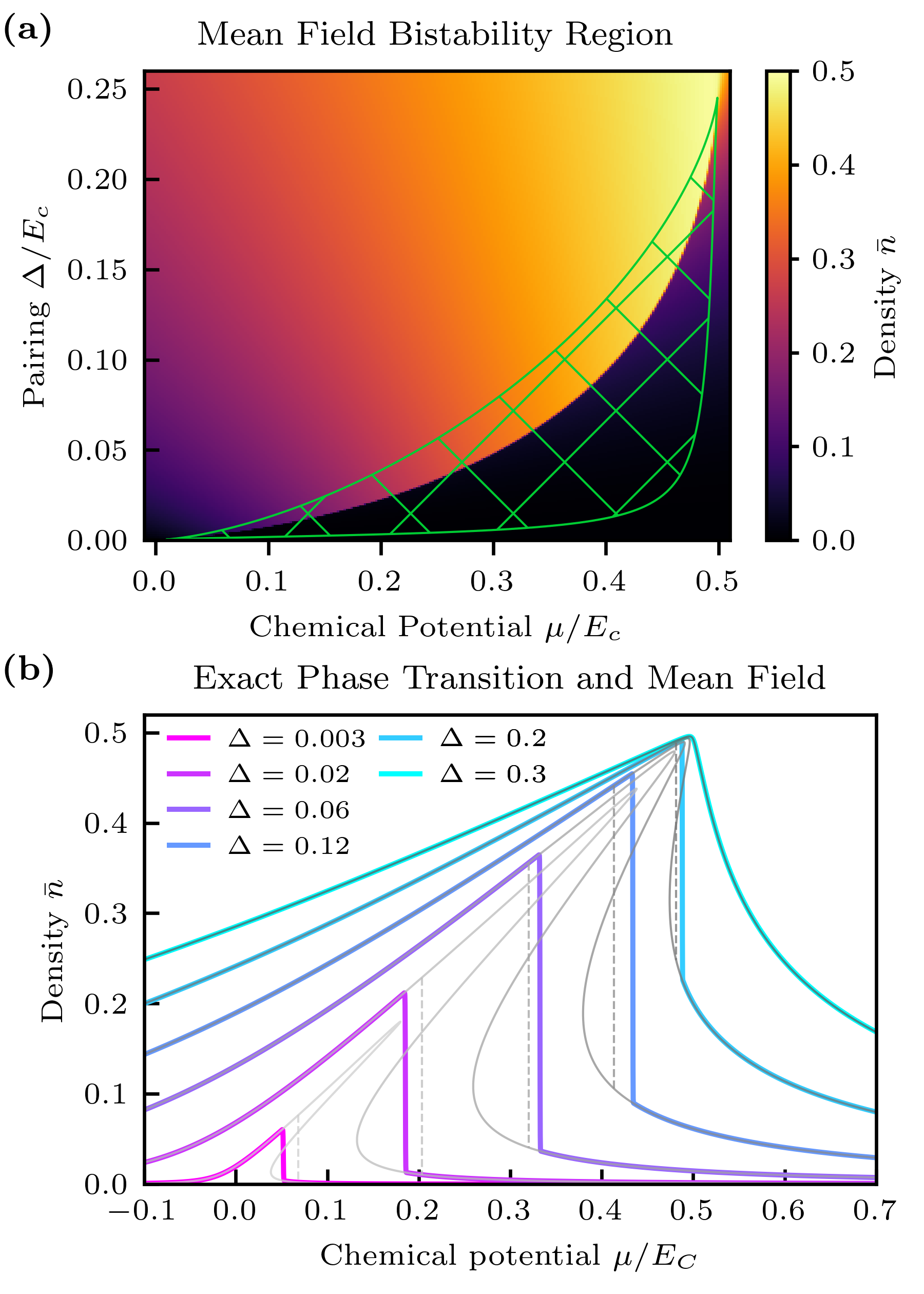}
	\caption{
	\textbf{Mean-field bistability of the particle density.}
    \textbf{(a)} The mean density $\bar{n}$ for $L=400$ sites, $E_C=1$ and $\kappa/E_C = 0.01$ (cf. Fig.~\ref{fig:phase-diagram}(a)) is shown in the $\mu$--$\Delta$ plane. The bistable region of Eq.~\eqref{eq:nbar-self-cons} is shown as the hatched region on the plot.
    \textbf{(b)} The exact solution particle density $\bar{n}$ is plotted as a function of chemical potential $\mu/E_C$ for the indicated pairing amplitudes $\Delta/E_C$, for parameters $L=2000$, $E_C = 1$, and $\kappa/E_C=0.01$
    The thin solid grey lines show the self-consistent solutions of Eq.~\eqref{eq:nbar-self-cons} for the same parameters.
    The vertical dashed lines indicate the expected transition point of the mean-field theory using the equal-area Maxwell construction.}
	\label{fig:mean-field}
 \end{figure}

Solving Eq.~\eqref{eq:nbar-self-cons} for $\bar{n}$ numerically (setting $E_C\equiv 1$), we find at least one valid (i.e., $0\leq \bar{n}\leq 0.5$) solution for all parameters.
In a particular region of the $\mu$--$\Delta$ plane, there are three solutions, indicating bistability of the mean-field theory.
This region is shown in Fig.~\ref{fig:mean-field}(a).
The bistable region exists only for $0\leq \mu \leq 0.5$, and it encompasses the first-order boundary line of the exact solution.
Moreover, as shown in Fig.~\ref{fig:mean-field}(b), the two stable branches of mean-field density predict the exact density well above and below the transition; however, a Maxwell equal-area construction for the transition point does not correctly predict the actual transition point.

\section{Mapping to a dissipative transverse-field Ising model}

Here, we show that the expectation value equations of motion of the mean-field fermionic system map to those of the mean-field theory of a transverse-field Ising model with an inhomogeneous transverse field under an Anderson pseudospin mapping.
Remarkably, this mapping holds despite any direct mapping of fermion loss to spin dissipation.

The fundamental excitations of the fermion system are Cooper pairs $\hat c_k^\dagger \hat c_{-k}^\dagger$ (cf.~Eq.~\eqref{eq:Bdagger}), which suggests we use the pseudospin mapping \cite{anderson_RandomPhase_1958}
\begin{align}
    \hat \sigma_k^{-} = \hat c_{-k}\hat c_k,\quad \hat \sigma_k^z = \hat n_k +\hat n_{-k} -1.
\end{align}
Here we restrict the momentum index $0\leq k < \pi$ to avoid double counting and let the $k=0$ and $k=\pi$ modes form a pair. Neither mode is excited by the Hamiltonian pairing so this pair always relaxes to vacuum.

Within the subspace of unbroken pairs, the mapping is algebraically faithful: this definition of the spin operators forms the spin-1/2 representation of SU(2).
Moreover, the momentum-space Hamiltonian (cf.~Eq.~\eqref{app-eq:momentum-H}) can be mapped exactly to a transverse field Ising model with inhomogeneous transverse fields:
\begin{align}
    \hat H &= -\mu\sum_{k} \hat \sigma_k^z + \frac{E_C}{2L}\sum_{k,k^\prime}(\hat\sigma_k^z +1) (\hat\sigma_{k^\prime}^z +1) + \sum_{k}2\Delta_k\hat\sigma_k^x,
\end{align}
where $\Delta_k = \Delta \sin k$ is the inhomogeneous Rabi drive strength.
Here, the index $k$ should be interpreted merely as a label for the $L/2$ spins which happens to coincide with momentum $\pm k$ modes in the fermionic system; no spatial structure is implied in the spin model.
It is clear that the mean-field Hamiltonian is
\begin{align}
    \hat H^{\rm (mf)} = \left( -\mu + \frac{E_C}{2}(\bar{M}+1)\right)\hat \sigma_k^z + \sum_{k}2\Delta_k\hat\sigma_k^x,
    \label{eq:spin-mf-H}
\end{align}
where $\bar{M} = (L/2)^{-1}\sum_k \langle\hat\sigma_k^z\rangle$ is the mean magnetization.
It is related to the mean fermionic density by $\bar{n} = (\bar{M}+1)/2$, which follows directly from the pseudospin mapping.

Although the single particle loss of the fermionic model breaks Cooper pairs, we find that for the noninteracting limit ($E_C=0$) and mean-field interaction (cf.~Eqs.~\eqref{eq:mf-int} and \eqref{eq:spin-mf-H}), there is a correspondence in the dynamics of expectation values
when fermion loss is mapped to the combination of spin loss and dephasing:
\begin{align}
    \kappa\Big(\mathcal{D}[\hat c_k]+\mathcal{D}[\hat c_{-k}]\Big)\hat\rho_{\rm fermi} \simeq \kappa\Big(\mathcal{D}[\hat \sigma_k^-] + \frac{1}{4}\mathcal{D}[\hat\sigma_k^z]\Big)\hat\rho_{\rm spin}.
    \label{eq:dissipation-equivalence}
\end{align}
Specifically, we show in App.~\ref{app:tfim} that given a fermion operator $\hat O_{\rm fermi}$ which maps to a spin operator $\hat O_{\rm spin}$, the equivalence
\begin{align}
    \langle\hat O_{\rm fermi}(t)\rangle = \langle\hat O_{\rm spin}(t)\rangle
\end{align}
holds when the fermion operator is evolved under the single particle loss dissipation and the spin operator is evolved under the spin loss and dephasing dissipation in Eq.~\eqref{eq:dissipation-equivalence}.

The intuition here is that single particle loss breaks a Cooper pair, thereby reducing the pair coherence, but it does not reduce particle density by as much as if a complete Cooper pair were lost.
Indeed, the particle density is reduced by half as much.
The result is an effective dephasing of the pairs in excess of that caused by excitation loss.
We stress that this is an effective mapping, not an exact mapping, as the fermion dissipation takes the system out of the subspace in which the pseudospin mapping holds.
Remarkably, despite the fermion dissipation taking the system out of the pseudospin subspace, this effective mapping holds for the mean-field theory.
However, as we show in App.~\ref{app:tfim}, the mapping breaks down for the full interacting Hamiltonian because the density-density interactions show sensitivity to whether or not the system is in the pseudospin subspace.

Because the mean-field fermionic model maps to a mean-field TFIM, one may expect the self-consistency equation for the particle density (cf.~Eq.~\eqref{eq:nbar-self-cons}) to take the usual form of a cubic polynomial in $\bar n$ \cite{roberts_Exact_2023,marcuzzi_Universal_2014,carr_Nonequilibrium_2013,lee_Antiferromagnetic_2011}.
However, this is clearly not the case, as even when rearranged to remove the square roots, Eq.~\eqref{eq:nbar-self-cons} is quartic in $\bar n$.
The reason for this is simply the inhomogeneity of the Rabi drive strengths: $\Delta_k = \Delta \sin k$, as shown in App.~\ref{app:mean-field}.
Were the drives uniform, Eq.~\eqref{eq:nbar-self-cons} would be the usual cubic polynomial in $\bar n$.


\section{Hidden time-reversal symmetry}

\begin{figure}[t!]
	\centering
	\includegraphics[width=3.375in]{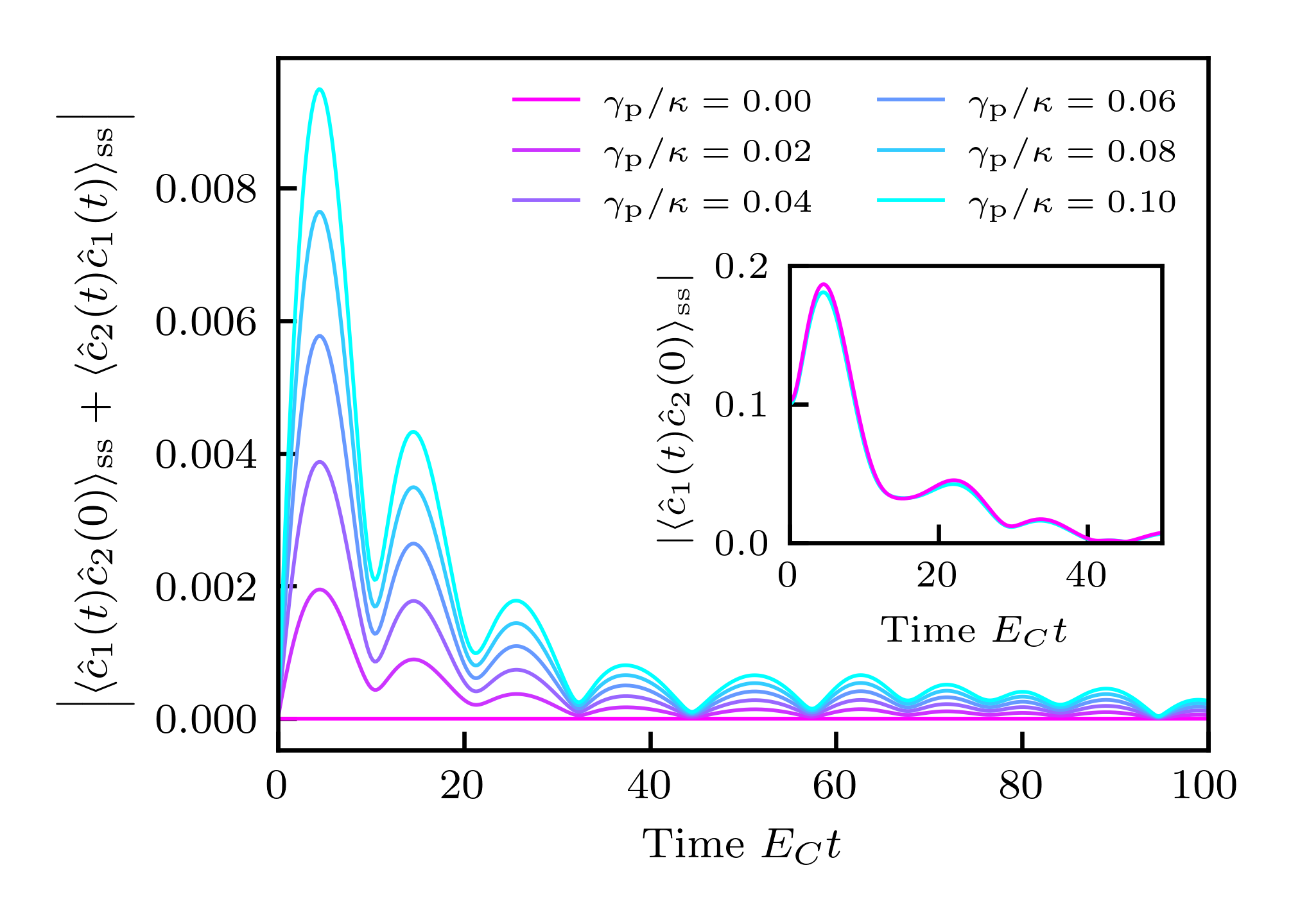}
	\caption{
	\textbf{Hidden time-reversal symmetry.} The magnitude of the difference between the ``forward'' correlation function $\langle \hat c_1(t)\hat c_2(0)\rangle_{\rm ss}$ and the ``reversed'' correlation function, $- \langle \hat c_2(t)\hat c_1(0)\rangle_{\rm ss}$ (note the minus sign) is plotted as a function of time $t$.
    Hidden time-reversal symmetry implies time-(anti)symmetry of this correlation function (cf.~Eq.~\eqref{eq:htrs-corr}). 
    Adding a small perturbation of weak pumping on site 1 with rate $\gamma_{\rm p}$ (cf.~Eq.~\eqref{eq:htrs-perturbation}) breaks hidden time-reversal symmetry, as evidenced by the nonzero difference between the forward and reversed correlation functions for $\gamma_{\rm p}>0$.
	\textbf{Inset:} The magnitude of the correlation function $\langle \hat c_1(t)\hat c_2(0)\rangle_{\rm ss}$ is plotted for $\gamma_{\rm p} = 0$ and $\gamma_{\rm p} = 0.1\kappa$.
	For both plots in this figure, the remaining parameters are $E_C \equiv 1$, $\mu = 0.2$, $\Delta = 0.15$, $\kappa = 0.01$, and $L=8$ sites.}
	\label{fig:htrs}
 \end{figure}

The existence of hidden time-reversal symmetry in a model implies that a certain class of two-time correlation functions obeys an Onsager symmetry \cite{roberts_Hidden_2021}. In particular, for bosonic and spin systems, steady state correlation functions of the effective non-hermitian Hamiltonian $\hat H_{\rm eff} = \hat H - \sum_j \hat L^\dagger_j \hat L_j$ and the dissipators $\hat L_j$ are time-symmetric.
Here, we show that the same Onsager symmetry holds in the fermionic model, with one slight modification: because the jump operators $\hat c_j$ anticommute, the two-time correlation functions of two jump operators is antisymmetric.
For example,
\begin{align}
	\langle \hat c_1(t)\hat c_2(0)\rangle_{\rm ss} = - \langle \hat c_2(t)\hat c_1(0)\rangle_{\rm ss}. \label{eq:htrs-corr}
\end{align}
The Onsager symmetry between $\hat H_{\rm eff}$ and any jump operator takes the standard form $\langle\hat H_{\rm eff}(t)\hat c_j(0)\rangle_{\rm ss} = \langle\hat c_j(t)\hat H_{\rm eff}(0)\rangle_{\rm ss}$, because it is bilinear in fermion ladder operators.

We demonstrate the existence of the Onsager symmetry Eq.~\eqref{eq:htrs-corr}, thus the existence of hTRS, and its subsequent breaking via a weak dissipative impurity added to the Lindbladian $\mathcal{L}$ (cf.~Eq.\eqref{eq:qme}).
Consider weak pumping on site 1: 
\begin{align}
    \mathcal{L}_{1}\hat\rho = \mathcal{L}\hat\rho + \gamma_{\rm p}\mathcal{D}[\hat c_1^\dagger]\hat\rho,
    \label{eq:htrs-perturbation}
\end{align}
where the pumping rate is $\gamma_{\rm p}\ll \kappa$.
As shown in Fig.~\ref{fig:htrs}, this small perturbation breaks the time (anti)symmetry of Eq.~\eqref{eq:htrs-corr}.
One also finds that a perfect absorber solution is broken by this perturbation.
Thus, not only does the CQA technique extend to fermionic models, so does the associated hidden time-reversal symmetry and its consequences.


\section{Conclusion}

We have introduced a class of interacting fermionic models whose dissipative steady states can be found exactly and which represent the first known fermionic models to possess hidden time-reversal symmetry.
These models exhibit first order phase transitions in the dissipative steady state, and and they have a class of time-symmetry two-time correlation functions, as predicted by hTRS.
Our work resolves the open questions of whether driven-dissipative fermionic models can possess hTRS, and, thus, whether steady state solutions of these models can be found using the coherent quantum absorber method.
We have explicitly shown that CQA may be applied to fermionic systems under fairly general circumstances.
By finding a class of solvable fermionic models, our work opens the possibility for a broader class of dissipative many-body systems, no longer restricted to bosonic or spin models, to be solvable via CQA.
We have also found a surprising connection between the fermionic system and a dissipative transverse-field Ising model which may find applications as approximation tool for solving interacting spin models by mapping them to solvable fermionic models.


\section*{Acknowledgments}

We thank David Noachtar for valuable discussions.
This work was supported by the Army Research Office under grant W911NF-25-1-0286, the Air Force Office of Scientific Research MURI program under Grant No. FA9550-19-1-0399, and the Simons Foundation through a Simons Investigator award (Grant No. 669487).

\appendix


\section{Coherent quantum absorber exact solution}
\label{app:cqa-general-soln}

Consider the most general model of $L$ spinless fermion sites discussed in the main text and given by Eqs.~\eqref{eq:H-general} and \eqref{eq:qme}.
Here we present the exact steady state solution $\mathcal{L}\hat\rho_{\rm ss}=0$ of this master equation for any antisymmetric pairing matrix $\Delta_{ij}$ and for all parameters $\kappa>0$, $\mu$, and $E_c$ in terms of the pure steady state wavefunction of the coherent quantum absorber.


\subsection{Coherent quantum absorber}

The coherent quantum absorber (CQA) system is constructed for Eq.~\eqref{eq:qme} by introducing an absorber with Hamiltonian $\hat H_B = -\hat H_A$ and cascading the dissipation:
\begin{align}
    \partial_t \hat\rho\cqa &= -i[\hat H\cqa,\hat \rho\cqa] + \sum_j \kappa\mathcal{D}[\hat c_{j,A} - \hat c_{j,B}]\hat\rho\cqa, \label{app-eq:cqa-qme}
    \\
    \hat H\cqa &= \hat H_A - \hat H_B - \frac{i\kappa}{2} \sum_j \left(\hat c_{j,A}^\dagger \hat c_{j,B} - {\rm H.c.} \right),\label{app-eq:H-cqa}
\end{align}
where the final term in $\hat H\cqa$ is the cascaded interaction.

Note that in general, the dissipators of the absorber $\hat L_{j,B}$ can be linear combinations of the system dissipators $\hat L_{j,A}$ \cite{roberts_Hidden_2021}:
\begin{align}
    \hat L_{j,B} = -\sum_k U_{jk}\hat L_{j,A},
\end{align}
where $U^\dagger U = U^2 = 1$ is an involution. Here we assume that $U=1$ such that $\hat L_{j,B} = -\hat c_{j,A}$, and we will find it to be correct.

We introduce the basis of collective ``dark'' and ``bright'' modes given by
\begin{align}
    \hat c_{j,+} &= \frac{1}{\sqrt 2}(\hat c_{j,A} + \hat c_{j,B}) \,\,:\,\, {\rm dark}, \label{app-eq:dark-mode}
    \\
    \hat c_{j,-} &= \frac{1}{\sqrt 2}(\hat c_{j,A} - \hat c_{j,B}) \,\,:\,\, {\rm bright},\label{app-eq:bright-mode}
\end{align}
respectively.
The dissipation of Eq.~\eqref{app-eq:cqa-qme} reduces to $\sum_j 2\kappa\mathcal{D}[\hat c_{j,-}]\hat\rho\cqa$, from which the bright/dark nomenclature becomes clear: the bright modes $\hat c_{j,-}$ decay due to the collective dissipation while the dark modes $\hat c_{j,+}$ do not decay; hence, they are dark to the dissipation.

The pure steady state wavefunction $|\psi\cqa\rangle$ of the coherent quantum absorber must satisfy the dark state conditions (cf.~Eq.~\eqref{eq:dark-cond}).
The jump operator dark state conditions imply that any excitations in the wavefunction must be in the dark modes only. 
Thus, we make the ansatz
\begin{align}
    |\psi\cqa\rangle = \sum_n \sum_{j_1,\dots,j_n} \alpha^{(n)}_{j_1,\dots,j_n} \hat c_{j_1,+}^\dagger \cdots \hat c_{j_n,+}^\dagger |0\rangle, \label{app-eq:general-cqa-ansatz}
\end{align}
where the $\alpha^{(n)}_{j_1,\dots,j_n}$ are the state vector coefficients and $|0\rangle$ is the fermion vacuum annihilated by all $\hat c_{j,\pm}$.
This wavefunction satisfies the dissipation dark state conditions by construction, and the state vector coefficients are determined by the Hamiltonian eigenstate condition.
Because $|\psi\cqa\rangle$ can contain only dark mode excitations, we can reduce $\hat H\cqa$ to the effective non-Hermitian Hamiltonian $\mathcal{\hat H}\cqa$ which excludes any terms $\propto \hat c_{j,-}$ that remove bright state excitations:
\begin{align}
    \mathcal{\hat H}\cqa &= -\sum_j \left( \tilde{\mu} - \frac{E_c}{2L}\hat N \right)\hat c_{j,-}^\dagger \hat c_{j,+} + \frac{1}{2}\sum_{ij}\Delta_{ij}\hat c_{i,-}^\dagger \hat c_{j,+}^\dagger \nonumber \\
    &= \mathcal{\hat H}_0 + \mathcal{\hat H}_\Delta, \label{app-eq:H-eff}
\end{align}
where $\tilde{\mu} = \mu + i\kappa/2$ is the complex chemical potential and $\hat N = \sum_{j,s} \hat n_{j,s}$ is the total number operator. 

The effective Hamiltonian divides into two parts: the particle-number-conserving $\mathcal{\hat H}_0$, whose action on $|\psi\cqa\rangle$ is to replace a dark mode excitation with a bright mode excitation, and the coherent pairing $\mathcal{\hat H}_{\Delta} \propto \Delta_{ij}$ which creates a pair of excitations, one in the dark mode subspace and one in the bright mode subspace. 
The Hamiltonian eigenstate condition $\hat H\cqa|\psi\cqa\rangle = 0$ is equivalent to
\begin{align}
    \mathcal{\hat H}\cqa|\psi\cqa\rangle = \mathcal{\hat H}_0|\psi\cqa\rangle + \mathcal{\hat H}_\Delta|\psi\cqa\rangle = 0. \label{app-eq:Heff-eigen-condition}
\end{align}
In general, the eigenstate condition requires only that $|\psi\cqa\rangle$ is an eigenstate of $\hat H\cqa$ \cite{roberts_Hidden_2021}; however, given that every term in $\mathcal{\hat H}\cqa$ creates a bright state excitation, only a zero-energy eigenstate is permitted.


\subsection{Wavefunction ansatz and the defect operator}

The most general ansatz for the CQA wavefunction is Eq.~\eqref{app-eq:general-cqa-ansatz} where $n$ indexes the number of fermions and $\alpha^{(n)}_{j_1,\dots,j_n}$ records the amplitude for those $n$ fermions to be on the specific sites $j_1,\dots,j_n$.
In its most general form, $|\psi\cqa\rangle$ could be a mixed parity state with all possible numbers of fermions $n=0,1,2,\dots$, but such a state would break the superselection rule \cite{wick_Intrinsic_1952}.
The effective Hamiltonian Eq.~\eqref{app-eq:H-eff} conserves total particle number parity (as does the full Hamiltonian $\hat H\cqa$), and for $\Delta_{ij} = 0$, the fermion vacuum $|\psi\cqa\rangle = |0\rangle$ is the unique steady state wavefunction.
Given these two facts and assuming continuity of the state as the pairing is turned on, we make the ansatz that in general $|\psi\cqa\rangle$ is an even parity state; thus, only $n=0,2,4,\dots$ terms are allowed in Eq.~\eqref{app-eq:general-cqa-ansatz}.

We introduce two pair creation operators: the ``dark-pair'' operator $\hat b_{ij}^\dagger$ and the ``defect operator'' $\hat \chi_{ij}^\dagger$ which are defined by
\begin{align}
    \hat b_{ij}^\dagger &\equiv \hat c_{i,+}^\dagger\hat c_{j,+}^\dagger,
    \\
    \hat \chi_{ij}^\dagger &\equiv \left( \hat c_{i,-}^\dagger \hat c_{j,+}^\dagger - \hat c_{j,-}^\dagger \hat c_{i,+}^\dagger \right).
\end{align}
Any even parity state in the dark subspace can be created from products of dark-pair creation operators $\hat b_{ij}^\dagger$.
The defect operators $\chi_{ij}^\dagger$ create bright-state-containing ``defects'' when they act on a state in the dark subspace.
Each operator is antisymmetric under index exchange, $\hat b_{ji}^\dagger  = -\hat b_{ij}^\dagger$ and $\hat \chi_{ji}^\dagger = \hat \chi_{ij}^\dagger$, and they behave like hard-core bosons: they commute and their products vanish for any shared index,
\begin{align}
    [\hat b_{ij}^\dagger,\hat b_{kl}^\dagger] &= [\hat\chi_{ij}^\dagger,\hat\chi_{kl}^\dagger] = [\hat b_{ij}^\dagger,\hat\chi_{kl}^\dagger] = 0,
    \\
    \hat b_{ij}^\dagger \hat b_{jk}^\dagger &= \hat \chi_{ij}^\dagger \hat \chi_{jk}^\dagger = \hat b_{ij}^\dagger \hat \chi_{jk}^\dagger  = 0.
\end{align}
We use this algebra to construct a state $|\psi\cqa\rangle$ for which the two terms in $\mathcal{\hat H}\cqa$ destructively interfere.

The pairing term $\mathcal{\hat H}_\Delta$ of the effective Hamiltonian (cf.~Eq.~\eqref{app-eq:H-eff}) can be written explicitly in terms of the defect operator:
\begin{align}
    \mathcal{\hat H}_\Delta &= \frac{\Delta}{2}\sum_{ij}M_{ij}\chi_{ij}^\dagger,
    \\
    M_{ij} &\equiv \frac{\Delta_{ij}}{||\Delta_{ij}||} = \frac{\Delta_{ij}}{\Delta},
\end{align}
where $M_{ij}$ is the normalized pairing matrix, $||M_{ij}||^2={\rm Tr}\,[M^\dagger M] = 1$, and $\Delta = ||\Delta_{ij}||$ is the effective pairing rate.

The particle-conserving Hamiltonian $\mathcal{\hat H}_0$ cannot be simply represented in terms of $\hat b_{ij}^\dagger$ and $\hat \chi_{ij}^\dagger$, but its commutator with the dark-pair operator can be. Specifically, 
\begin{align}
    [\mathcal{\hat H}_0,\hat b_{ij}^\dagger] = - \hat \chi_{ij}^\dagger \left[ \tilde{\mu} - \frac{E_{c}}{2L} \big(\hat{N}+2\big) \right] + \frac{E_{c}}{L}\hat{b}_{ij}^{\dagger}\sum_{n}\hat{c}_{n,-}^{\dagger}\hat{c}_{n,+}.
\end{align}
The first term in this commutator $\propto \hat\chi_{ij}^\dagger$ is the desired algebraic relation that makes the defect and dark-pair operator picture useful. The second term $\propto \hat b_{ij}^\dagger$ is not immediately in a useful form, but notice that it annihilates the fermion vacuum $|0\rangle$ and its commutator with a second $\hat b_{kl}^\dagger$ is simply
\begin{align}
    \frac{2E_c}{L} \hat b_{ij}^\dagger \sum_n [\hat c_{n,-}^\dagger\hat c_{n,+},\hat b_{kl}^\dagger] = \frac{E_c}{L} \hat\chi_{kl}^\dagger \hat b_{ij}^\dagger.
\end{align}
Since every term in the commutator $[\mathcal{\hat H}_0,\hat b_{ij}^\dagger]$ ultimately results in a defect operator, we have a systematic way to match the action of $\mathcal{\hat H}_0$ on a particular configuration of dark pairs to the action of $\mathcal{\hat H}_\Delta$ on another configuration.


\subsection{General exact solution}

Consider the following ansatz for $|\psi\cqa\rangle$ consisting of all possible number of dark pairs delocalized across the system according to the pairing matrix $M_{ij} = \Delta_{ij}/\Delta$:
\begin{align}
    |\psi\cqa\rangle &= \sum_{n=0}^{\infty} \frac{\alpha_n}{n!} \big( \hat B^\dagger \big)^n|0\rangle, \label{app-eq:cqa-ansatz}
    \quad
    \hat B^\dagger \equiv \frac{1}{2}\sum_{ij}M_{ij} \hat b_{ij}^\dagger 
\end{align}
Here $\alpha_n$ is the state vector coefficient for the state with $n$ dark pairs and $\hat B^\dagger$ is the delocalized dark-pair creation operator.
The factor of $1/2$ in $\hat B^\dagger$ accounts for the double counting $M_{ji}\hat b_{ji}^\dagger = M_{ij}\hat b_{ij}^\dagger$.
We claim that there exists a unique set of $\alpha_n$, with $\alpha_0= 1$, for which this is a zero-energy eigenstate of $\mathcal{\hat H}\cqa$ (cf.~Eq.~\eqref{app-eq:H-eff}).

To find $\alpha_n$, we impose the zero-energy eigenstate condition: $\mathcal{\hat H}\cqa|\psi\cqa\rangle = 0$.
Commuting $\mathcal{\hat H}_0$ past powers of $\hat B^\dagger$, one finds
\begin{align}
    [\mathcal{\hat H}_0,(\hat B^\dagger)^n] = -n\left( \tilde{\mu} - \frac{n}{L} E_c \right)\sum_{ij} \frac{1}{2}M_{ij}\hat\chi_{ij}^\dagger (\hat B^\dagger)^{n-1}.
\end{align}
The pairing Hamiltonian $\mathcal{\hat H}_\Delta$ simply creates a defect on any existing configuration of dark state pairs. Thus, the zero-energy eigenstate condition reads
\begin{align}
    \sum_n \frac{1}{n!} &\left[ -\left(\tilde{\mu} - \frac{n+1}{L}E_c\right)\alpha_{n+1} +  \Delta\alpha_n\right]
    \\
    &\qquad\quad \times\,\,\Bigg( \sum_{ij} \frac{1}{2}M_{ij}\hat\chi_{ij}^\dagger (\hat B^\dagger)^{n}|0\rangle \Bigg) = 0,
    \nonumber
\end{align}
which must vanish separately for each nonzero defect $M_{ij}\hat\chi_{ij}^\dagger$. 
Therefore, we obtain a single-term recurrence relation for the $\alpha_n$:
\begin{align}
    \alpha_n = \frac{\Delta}{\tilde{\mu} - \frac{n}{L}E_c} \alpha_{n-1}.
    \label{app-eq:general-recurrence-alpha}
\end{align}
Taking the initial condition $\alpha_0 = 1$, the solution is
\begin{align}
    \alpha_n = \frac{\Delta^n}{\prod_{m=1}^n [\tilde{\mu} - \frac{m}{L}E_c]}. \label{app-eq:alpha-n-solution}
\end{align}
Thus, we have the general form of the exact steady state wavefunction $|\psi\cqa\rangle$ for any pairing matrix $\Delta_{ij}$, up to normalization.
In general normalization is a nontrivial task, as computing the component state norms $||(\hat B^\dagger)^n||^2$ requires enumerating all configurations of $n$ non-overlapping pairs of fermions, weighted by $\Delta_{ij}$.


\section{Cascaded fermionic systems}
\label{app:cascaded-fermionic-systems}

The theory of cascaded quantum systems allows for the construction of nonreciprocal quantum systems at the expense of introducing dissipation \cite{carmichael_Quantum_1993,gardiner_Driving_1993}.
Under very general assumptions, interactions between otherwise isolated systems $A$ and $B$ of the form
\begin{align}
    \hat H_{\rm int} = \lambda(\hat A \hat B + {\rm H.c.}) \label{app-eq:Hint-general}
\end{align}
can be cascaded for spin and bosonic systems with arbitrary operators $\hat A$ and $\hat B$, and for fermionic systems when $\hat A$ 
and $\hat B$ comprise even numbers of fermionic creation and annihilation operators of the respective systems \cite{metelmann_Nonreciprocal_2017}.
This class of fermionic interactions \emph{does not include}, for example, tunneling between two systems, $\hat H_{\rm int} = -t(\hat c_{i,A}^\dagger\hat c_{j,B} + {\rm H.c.})$.
However, we show that a slight modification of the general formulation allows for a broad class of fermionic systems to be cascaded, in the sense that \emph{observables} of one system are influenced by the other but not vice versa.


\subsection{Nonreciprocity in quantum systems}

Nonreciprocity in spin and bosonic systems is easy to characterize: a nonreciprocal interaction between two isolated systems $A$ and $B$ in one for which system $A$ influences system $B$ but not vice versa, as measured by the equations of motion for system-local operators (e.g., $\hat X_A = (\hat X)_A\otimes (\hat 1)_B$).
If the equations of motion for all $\langle\hat X_A\rangle$ have no dependence on system $B$ quantities, but there is some $\langle \hat X_B \rangle$ whose EOM depends on system $A$, then we have $A\to B$ nonreciprocity; system $A$ is said to be upstream of system $B$.

Unlike spin or bosonic composite systems, which are formed under tensor products of the subsystem Hilbert spaces, composite fermionic systems are formed under the exterior product, which means that truly ``isolated'' fermionic subsystems do not exist.
As we will show, this implies that for the class nonreciprocal fermion interactions (cf.~Eq.~\eqref{app-eq:Hint-general}) where $\hat A$ and $\hat B$ are odd in fermionic creation and annihilation operators (i.e., comprise odd powers of ladder operators), nonreciprocity of the equations of motion is only guaranteed for local operators $\hat X_{A/B}^{\rm even}$ that are \emph{even} in fermionic creation and annihilation operators.

For physical fermionic master equations (i.e., those that could be realized with actual indistinguishable fermions), the density matrix must obey the parity superselection rule, which precludes superpositions between states with even numbers of fermions and states with odd numbers of fermions \cite{wick_Intrinsic_1952}.
The imposition of superselection enforces that odd operator expectation values are zero, $\langle\hat X^{\rm odd}\rangle(t) = 0$.
Therefore, as we will show, any such fermionic system can be cascaded.


\subsection{The standard formulation}

The general standard formulation to engineer a noreciprocal interaction is as follows \cite{metelmann_Nonreciprocal_2017}:
Given two quantum systems $A$ and $B$ that have no existing interactions or couplings between them, we seek to make the interaction Eq.~\eqref{app-eq:Hint-general} between the systems nonreciprocal
First, we assume that $[\hat A,\hat X_B] = [\hat B,\hat X_A] = [\hat A,\hat B] = 0$ for all operators $\hat X_i$ of system $i$. 
For spins or bosons, these assumptions are automatically satisfied if the two systems are isolated. 
For fermions, this requires only that $\hat A$ and $\hat B$ are even in fermionic creation and annihilation operators (see e.g., Ref.~\cite{malz_Current_2018} for such an example).
We introduce the interaction $\hat H_{\rm int}$ and a Markovian reservoir that couples the two systems via $\hat A$ and $\hat B$:
\begin{align}
    \partial_t \hat\rho = -i[\hat H_A + \hat H_B + \hat H_{\rm int},\hat\rho] + \kappa\mathcal{D}[\hat A + e^{i\phi}\hat B^\dagger]\hat\rho, \label{app-eq:nonrecip-qme}
\end{align}
where $\hat H_{\rm int}$ is given in Eq.~\eqref{app-eq:Hint-general},  $\kappa$ is the dissipation rate, and $\phi$ is a relative phase that cannot be gauged away.
By choosing the dissipation rate $\kappa$ and the phase angle $\phi$ correctly, we obtain nonreciprocity in the equations of motion for all system-local operators $\hat X_i$.

Explicitly, the equations of motion for a arbitrary system-local operators $\hat X_A$ and $\hat X_B$ are
\begin{align}
    \partial_t \langle\hat X_A\rangle &= -i\langle[\hat X_A,\hat H_A]\rangle + \kappa\langle\mathcal{D}^\dagger[\hat A]\hat X_A\rangle 
    \label{app-eq:A-eom-commuting}
    \\
    &-i\Big(\lambda -\frac{i\kappa}{2} e^{i\phi} \Big)^*\langle[\hat X_A,\hat A]\hat B\rangle 
    \nonumber
    \\
    &-i\Big(\lambda - \frac{i\kappa}{2} e^{i\phi} \Big)\langle\hat B^\dagger[\hat X_A,\hat A^\dagger]\rangle,
    \nonumber
    \\
    \partial_t \langle\hat X_B\rangle &= -i\langle[\hat X_B,\hat H_B]\rangle + \kappa\langle\mathcal{D}^\dagger[\hat B^\dagger]\hat X_B\rangle 
    \label{app-eq:B-eom-commuting}
    \\
    &-i\Big(\lambda +\frac{i\kappa}{2} e^{i\phi} \Big)^*\langle\hat A[\hat X_B,\hat B]\rangle 
    \nonumber
    \\
    &-i\Big(\lambda +\frac{i\kappa}{2} e^{i\phi} \Big) \langle[\hat X_B,\hat B^\dagger]\hat A^\dagger\rangle.
    \nonumber
\end{align}
We therefore see that when the dissipation rate $\kappa$ and phase $\phi$ are tuned to 
\begin{align}
    \kappa \overset{!}{=} 2\lambda,\quad \phi \overset{!}{=} -\pi/2,
\end{align}
the equation of motion for every $\langle\hat X_A\rangle$ becomes independent of system $B$, while the EOM for $\langle\hat X_B\rangle$ can still depend on system $A$. (For $\phi=+\pi/2$, the nonreciprocity goes in the other direction with system $B$ influencing system $A$.)


\subsection{Nonreciprocal odd fermionic interactions}
\label{sec:lin-fermion-int}

Suppose now that we seek the nonreciprocal interaction Eq.~\eqref{app-eq:Hint-general} between two fermionic systems $A$ and $B$ for which $\hat A$ and $\hat B$ are odd in fermionic ladder operators.
For example, consider a nonreciprocal tunneling $\hat H_{\rm int} = \lambda(\hat c_{i,B}^\dagger\hat c_{j,A} + {\rm H.c.})$ where fermions can tunnel only from system $A$ to system $B$. 
Clearly, the assumption $[\hat A,\hat B]=0$ does not hold, as instead we have $\{\hat A,\hat B\}=0$.
This condition, along with $[\hat A,\hat X_B] = [\hat B,\hat X_A] = 0$, is sufficient to obtain nonreciprocity of \emph{observable system-local fermionic quantities} $\hat O_i$ (i.e., even operators $\hat O_i$) with only a slight tweak to the standard formulation.

We consider the same engineered reservoir setup as in Eq.~\eqref{app-eq:nonrecip-qme}. 
Again, $\kappa$ is the dissipation rate and $\phi$ is the phase that controls the direction of nonreciprocity.
The equations of motion for even fermionic observables $\hat O_A$ and $\hat O_B$ are
\begin{align}
    \partial_t\langle\hat O_A\rangle &= -i\langle[\hat O_A,\hat H_A]\rangle + \kappa\langle\mathcal{D}^\dagger[\hat A]\hat O_A\rangle 
    \\
    &-i\Big( \lambda + \frac{i\kappa}{2}e^{i\phi} \Big)^*\langle[\hat O_A,\hat A]\hat B\rangle 
    \nonumber
    \\
    &-i\Big( \lambda +\frac{i\kappa}{2}e^{i\phi} \Big)\langle\hat B^\dagger[\hat O_A,\hat A^\dagger]\rangle,
    \nonumber
    \\
    \partial_t\langle\hat O_B\rangle &= -i\langle[\hat O_B,\hat H_B]\rangle + \kappa\langle\mathcal{D}^\dagger[\hat B^\dagger]\hat O_B\rangle 
    \\
    &-i\Big( \lambda - \frac{i\kappa}{2}e^{i\phi} \Big)^*\langle\hat A[\hat O_B,\hat B]\rangle 
    \nonumber
    \\
    &-i\Big( \lambda - \frac{i\kappa}{2}e^{i\phi} \Big)\langle\hat [\hat O_B,\hat B^\dagger]A^\dagger\rangle.
    \nonumber
\end{align}
Comparing these equations of motion with Eqs.~\eqref{app-eq:A-eom-commuting} and \eqref{app-eq:B-eom-commuting}, we see that each $\propto i\kappa$ term has the opposite sign. Otherwise, the expressions are identical. This immediately tells us that for the dissipation rate and phase tuning
\begin{align}
    \kappa \overset{!}{=} 2\lambda,\quad \phi \overset{!}{=} +\pi/2,
\end{align}
system $A$ influences system $B$ but not vice versa. 
In a similar fashion as before, when the phase is changed to $\phi = -\pi/2$, the nonreciprocity flips direction.

The requirement that the commutators $[\hat A,\hat O_B] = [\hat B,\hat O_A] = 0$ imposes a single constraint on the Hamiltonians $\hat H_A$ and $\hat H_B$: they must be constructed from bilinear fermion operators. 
Thus, a Hamiltonian term such as the linear drive $\hat H_{\rm lin} = \Omega({\hat c + \hat c^\dagger})$ is not permitted. 
For physical fermionic systems which must obey parity superselection this places no real constraint; only when considering spin systems whose mapping to fermions might violate superselection could this be a problem.

To conclude, we note that for the CQA problem (cf.~Eq.~\eqref{eq:cqa-qme} in the main text), the collective loss dissipation is $\hat L_{j,{\rm cqa}} = \hat c_{j,A} - \hat c_{j,B}$. Thus, $\hat A_j = \hat c_{j,A}$ and $e^{i\phi}\hat B_j^\dagger = -\hat c_{j,B}$ (cf.~Eq.~\eqref{app-eq:nonrecip-qme}).
To achieve nonreciprocity from system $A$ to system $B$ we must set $\lambda=\kappa/2$ and $\phi=+\pi/2$ in the interaction Hamiltonian Eq.~\eqref{app-eq:Hint-general}, thus,
\begin{align}
    \hat H_{\rm int} &= \frac{\kappa}{2}\sum_j\left(-i \hat c_{j,A}\hat c_{j,B} + i\hat c_{j,B} \hat c_{j,A}^\dagger \right)
    \nonumber
    \\
    &= -\frac{i\kappa}{2}\sum_j\left(\hat c_{j,A}^\dagger \hat c_{j,B} - {\rm H.c.} \right).
\end{align}
Therefore, despite the gauge phase $\phi$ being opposite for the cascaded fermionic system, due to fermionic anticommutation relations, the overall sign of the cascaded interaction $\hat H_{\rm int}$ of the CQA system is the same as it is for spin and bosonic systems (see, e.g., Ref.~\cite{roberts_Hidden_2021}).


\section{Nearest neighbor pairing in 1D}
\label{app:1d-p-wave}

The concrete model studied in the main text has nearest neighbor pairing in 1D, with a Hamiltonian given by Eq.~\eqref{eq:H-1D}.
Although we only considered periodic boundary conditions (PBC) in the main text, here we present the solution for both PBC and open boundary conditions (OBC).
We find the exact steady state wavefunction (cf.~Eqs.~\eqref{app-eq:cqa-ansatz} and \eqref{app-eq:alpha-n-solution})
\begin{align}
    |\psi\cqa\rangle &= \sum_{n=0}^{\infty} \frac{\alpha_n}{n!} \big( \hat B^\dagger \big)^n|0\rangle,
    \label{app-eq:wavefunc-1d}
    \quad
    \hat B^\dagger = \sum_{j} \hat c_{j,+}^\dagger \hat c_{j+1,+}^\dagger,
    \\
    \alpha_n &= \frac{\Delta^n}{\prod_{m=1}^n [\tilde{\mu} - \frac{m}{L}E_c]},
    \label{app-eq:alpha-n-1d}
\end{align}
where $\tilde{\mu} = \mu + i\kappa/2$ is the complex chemical potential, and where the boundary conditions are implicitly included in $\hat B^\dagger$.
The sum over $j$ in $\hat B^\dagger$ terminates at $j_{\rm max}=L-1$ for OBC and at $j_{\rm max}=L$ for PBC where $L+1\equiv 1$. 

For all finite $L<\infty$ the sum over $n$ is finite as only ${\sim L/2}$ pairs can fit on the length $L$ chain.
Specifically, for OBC, the sum terminates at $n_{\rm max} = \lfloor L/2\rfloor$, and for PBC, the sum terminates at $n_{\rm max} = \lfloor (L-1)/2 \rfloor$.
A length $L=2m$ PBC chain \emph{cannot} be maximally filled because there are two distinct ways to construct the maximally-filled state from the nearest neighbor pairs.
These two configurations differ only by the relative position of $\hat c_{1}^\dagger$:
\begin{align}
    \frac{1}{(L/2)!}&(\hat B^\dagger)^{L/2}|0\rangle
    \\
    &= \frac{1}{2}\left[ (\hat c_1^\dagger\hat c_2^\dagger\cdots\hat c_{L-1}^\dagger\hat c_{L}^\dagger) + (\hat c_2^\dagger\hat c_3^\dagger\cdots\hat c_{L}^\dagger\hat c_{1}^\dagger) \right]|0\rangle.
    \nonumber
\end{align}
These two terms are related by the commutation of $\hat c_1^\dagger$ past $(L-1)$ fermion operators, hence they differ by a sign and perfectly cancel. Note that this conclusion is apparent in momentum space. The pairing dispersion $\Delta_k = \Delta\sin k$ vanishes for $k=0$, the maximally-filled state.


\subsection{Normalization: the combinatorics of dimers on a chain}

To compute observables and correlation functions of the steady state, the steady state must be normalized:
\begin{align}
    \mathcal{N}\cqa &\equiv \langle\psi\cqa|\psi\cqa\rangle = \sum_{n=0}^\infty |\alpha_n|^2 \mathcal{N}(L,n),
    \label{app-eq:wavefunc-norm}
    \\
    \mathcal{N}(L,n) &= ||\frac{1}{n!}(\hat B^\dagger)^n|0\rangle||^2 = \frac{1}{(n!)^2} \langle0|(\hat B)^n(\hat B^\dagger)^n|0\rangle,
    \nonumber
\end{align}
where $\mathcal{N}(L,n)$ is the state norm of the delocalized $n$-pair state. 

The problem of computing the state norms $\mathcal{N}(L,n) = ||\frac{1}{n!}(\hat B^\dagger)^n|0\rangle||^2$ for OBC (PBC) chains is equivalent to counting the number of unique ways that $n$ dimers can be placed on OBC (PBC) chains of $L$ sites without overlap.
For open boundaries we can give a simple combinatorial argument, then the solution for PBC follows almost trivially.
We note also that this problem can be generally solved in terms of a generating function \cite{fisher_Association_1960,fisher_Statistical_1961}, but this solution is not readily put into closed form.
Indeed, one may show by direct computation of the generating function that $\mathcal{N}_{\rm OBC}(L,n) = 2^{2n-L} \sum_{k=n}^{\lfloor L/2\rfloor} \binom{L+1}{2k+1}\binom{k}{k-n}$. The curious may consult App.~\ref{app:gen-func}.

Consider an OBC chain of $L$ sites onto which we will place $n \leq \lfloor L/2\rfloor $ non-overlapping dimers. Here, a dimer spans two adjacent sites on the chain. 
Let the number of unique configurations for a given $L$ and $n$ be denoted $\mathcal{N}_{\rm OBC}(L,n)$.
Each valid configuration of dimers on the chain can be viewed as an arrangement of $(L-n)$ objects on a line; namely, the $n$ dimers and $(L-2n)$ empty sites. The number of such arrangements is simply the binomial coefficient
\begin{align}
    \mathcal{N}_{\rm OBC} = \binom{L-n}{n}. \label{app-eq:obc-state-norm}
\end{align}

Now consider the chain with periodic boundaries. Notice that there are $L$ distinct places that the first dimer can be placed on the chain. 
Then, what remains is a length $(L-2)$ open boundary chain on which the remaining $(n-1)$ dimers must be placed.
Accounting for the $n$ different dimers which can be considered the first, the number of configurations of $n$ dimers on the PBC chain of length $L$ (and thus the PBC state norm of $n$ pairs) is
\begin{align}
    \mathcal{N}_{\rm PBC}(L,n) &= \frac{L}{n} \mathcal{N}_{\rm OBC}(L-2,n-1)
    \label{app-eq:pbc-state-norm}
    \\
    &= \frac{L}{n}\binom{L-n-1}{n-1}
\end{align}
for all $n>0$.
With these expressions for $\mathcal{N}(L,n)$, we may compute the normalization $\mathcal{N}\cqa = \langle\psi\cqa|\psi\cqa\rangle$ of the CQA wavefunction (cf.~Eq.~\eqref{app-eq:wavefunc-norm}).

For the remainder of this appendix, we will assume a normalized wavefunction $|\psi\cqa\rangle$, and assume an overall factor of $\mathcal{N}^{-1}\cqa$ implicitly appears in each of expectation values given below.


\subsection{Global expectation values}

Global expectation values, such as total particle number $\langle\hat N_A\rangle = \sum_j \langle\hat n_{j,A}\rangle$ (of the physical system $A$) or pairing operator $\langle \hat B^\dagger\rangle$ (of the CQA system), are readily computed using their algebraic properties and the form of the wavefunction. 
First note that since the wavefunction only contains dark mode excitations, any expectation value of a physical system operator is proportional to the equivalent dark mode operator. For example,
\begin{align}
    \langle \hat n_{j,A}\rangle &= \frac{1}{2}\left[\langle \hat n_{j,+}\rangle+\langle \hat n_{j,-}\rangle+\langle \hat c_{j,+}^\dagger c_{j,-}\rangle+\langle \hat c_{j,-}^\dagger c_{j,+}\rangle \right] \nonumber
    \\
    &= \frac{1}{2}\langle \hat n_{j,+}\rangle,
\end{align}
using the definitions of the bright and dark modes given by Eqs.~\eqref{app-eq:dark-mode} and \eqref{app-eq:bright-mode}.

The expectation value $\langle\hat B^\dagger\rangle$ is extremely easy to compute from the CQA wavefunction: its expectation value is simply the overlap of each term $\propto (\hat B^\dagger)^n$ in $\hat B^\dagger|\psi\cqa\rangle$ with the corresponding term $\propto (\hat B)^n$ in $\langle \psi\cqa|$, hence
\begin{align}
    \langle \hat B^\dagger\rangle &= \sum_{n=1} (n)\alpha_{n}^*\alpha_{n-1} \mathcal{N}(L,n),
    \label{app-eq:pair-corr-expr}
\end{align}
where $\mathcal{N}(L,n)$ is the boundary-condition-dependent state norm for $n$ pairs on $L$ sites (cf.~Eqs.~\eqref{app-eq:obc-state-norm} and \eqref{app-eq:pbc-state-norm}).
More generally, the expectation of powers of $\hat B^\dagger$ are readily found to be
\begin{align}
    \langle (\hat B^\dagger)^m\rangle = \sum_{n=m}\frac{n!}{(n-m)!}\alpha^*_n\alpha_{n-m}\mathcal{N}(L,n).
\end{align}
This sum terminate at $n_{\rm max} = \lfloor L/2\rfloor$ for OBC and $n_{\rm max} = \lfloor {(L-1)}/2 \rfloor$) for PBC.

The total particle number $\langle \hat N\rangle = \sum_j \langle\hat n_{j,+}\rangle$ of the CQA system is also readily computed. The commutator $[\hat N,\hat B^\dagger]= 2\hat B^\dagger$ implies that $[\hat N,(\hat B^\dagger)^n]= 2n (\hat B^\dagger)^n$. Thus, we find
\begin{align}
    \langle\hat N\rangle &= \sum_n (2n)|\alpha_n|^2 \mathcal{N}(L,n),
    \label{app-eq:Ntotal-expr}
\end{align}
This sum terminate at $n_{\rm max} = \lfloor L/2\rfloor$ for OBC and $n_{\rm max} = \lfloor {(L-1)}/2 \rfloor$) for PBC.
Furthermore, powers of the number operator are $\langle \hat N^m\rangle = \sum_n (2n)^m|\alpha_n|^2\mathcal{N}(L,n)$.


\subsection{Anomalous correlation functions}

The anomalous correlation functions are the expectation values
\begin{align}
    \langle\hat c_{i} \hat c_{j}\rangle \equiv \langle\psi\cqa|\hat c_{i,+} \hat c_{j,+}|\psi\cqa\rangle
    \label{app-eq:anom-corr-def}
\end{align}
for any pair of dark mode sites $i$ and $j$. The physical system correlations are scaled: $\frac{1}{2}\langle\hat c_{i}\hat c_{j}\rangle$.
For the arguments that follow, it is more physically transparent to consider the conjugate anomalous correlations
$\langle\hat c_i^\dagger \hat c_j^\dagger\rangle$ rather than the traditional expression $\langle\hat c_j \hat c_i\rangle$.
In all that follows, we assume $j>i$ without loss of generality and we suppress the dark-mode index on $\hat c_j \equiv \hat c_{j,+}$.

The correlation $\langle\hat c_i^\dagger \hat c_j^\dagger\rangle$ may be nonzero only if the state $\hat c_i^\dagger\hat c_j^\dagger|\psi\cqa\rangle$ has an overlap with $\langle\psi\cqa|$.
Due to the nearest-neighbor pairing, this simple fact implies that only those components of the wavefunction $|\psi\cqa\rangle$ which have a contiguous ``string'' of occupied sites \emph{between} site $i$ and site $j$ -- and with the boundary sites $i$ and $j$ empty -- contribute to $\langle\hat c_i^\dagger \hat c_j^\dagger\rangle$. For example, a wavefunction component of the form 
\begin{align}
    (\cdots\hat 1_i\cdot \hat c_{i+1}^\dagger\hat c_{i+2}^\dagger\cdots\hat c_{j-2}^\dagger\hat c_{j-1}^\dagger \cdot \hat 1_j \cdots)|0\rangle,
\end{align}
contributes to the correlation.
Here we explicitly write the identity $\hat 1$ for sites $i$ and $j$ to emphasize they must be unoccupied.
Moreover, such strings of occupied sites \emph{must have even length} because nearest-neighbor pairs can create only even-length occupied strings.
Then acting with $\hat c_i^\dagger \hat c_j^\dagger$ on such a string creates a new contiguous string comprising nearest-neighbor pairs and, therefore, has overlap with components of $\langle\psi\cqa|$.
Every component of $\hat c_i^\dagger \hat c_j^\dagger|\psi\cqa\rangle$ without a contiguous string of occupied sites spanning $[i,j]$ will necessarily have strings with odd length; these have no overlap with $\langle\psi\cqa|$.

For all OBC chains and even-length PBC chains, this constraint implies that the only nonzero anomalous correlations are of the form $\langle \hat c_j^\dagger \hat c_{j+2m+1}^\dagger\rangle$.
E.g., $\langle\hat c_1^\dagger\hat c_2^\dagger\rangle\neq 0$ and $\langle\hat c_1^\dagger\hat c_4^\dagger\rangle\neq 0$ but $\langle\hat c_1^\dagger\hat c_3^\dagger\rangle = 0$.
To see why, note that these chains can be decomposed into $A$ and $B$ sublattices, with all odd sites in sublattice $A$ and even sites in sublattice $B$.
Nearest neighbor pairs always span the sublattices, hence any contiguous string must terminate on the left on sublattice $A$ and on the right on sublattice $B$, or vice versa. 
Therefore, only anomalous correlations between sublattices are nonzero.
Odd-length PBC chains are different, however, as they have no decomposition into distinct sublattices.
Thus, anomalous correlations are generically nonzero between any two sites.


\subsection{Anomalous correlations in even-length PBC chains}

Here, we compute Eq.~\eqref{app-eq:anom-corr-def} for even-length PBC chains, as this is relevant to the discussion in Sec.~\ref{sec:phase-transition}.
The basic arguments employed here carry over to OBC chains and odd-length PBC chains with minor modifications.

Due to the translational invariance of the PBC chain and the fact that only cross-sublattice anomalous correlations are nonzero, it is sufficient to consider correlation functions of the form $\langle\hat c_1^\dagger \hat c_{2m}^\dagger\rangle$, for $1 \leq m\leq L/2$.
Using the form of $|\psi\cqa\rangle$ given in Eq.~\eqref{app-eq:wavefunc-1d}, it is clear the correlation function must be
\begin{align}
    \langle0|(\hat B)^{n} \hat c_1 ^\dagger \hat c_{2m}^\dagger (B^\dagger)^{n^\prime}|0\rangle \propto \delta_{n,n^\prime-1},
\end{align}
reducing the correlation function to a single sum over $n$.
Furthermore, the explicit appearance of $\hat c_1^\dagger$ and $\hat c_{2m}^\dagger$ imply that we can decompose $\hat B^\dagger$ as 
\begin{align}
    \hat B^\dagger &= \sum_{j=2}^{2m-2} \hat c_j^\dagger\hat c_{j+1}^\dagger + \sum_{j=2m+1}^{L-1}\hat c_j^\dagger\hat c_{j+1}^\dagger 
    \\
    &\equiv \hat B^\dagger_{L_1} + \hat B^{\dagger}_{L_2},
    \nonumber
\end{align}
where the sites $1$ and $2m$ are explicitly excluded. Here $L_1 = 2m-2$ is the length of the effective OBC chain spanning sites $[2,2m-1]$ and $L_2 = L-2m$ is the length of the OBC chain spanning sites $[2m+1,L]$.
The anomalous correlation is therefore
\begin{widetext}
\begin{align}
    \langle\hat c_1^\dagger\hat c_{2m}^\dagger\rangle 
    &=  \sum_{n=1}^{L/2-1}\frac{\alpha_n^*\alpha_{n-1}}{n!(n-1)!} \langle0| (\hat B)^n  (\hat c_1^\dagger\hat c_{2m}^\dagger) (\hat B_{L_1}^\dagger + \hat B_{L_2}^\dagger)^{n-1}|0\rangle
    \\
    &=  \sum_{n=1}^{L/2-1}\frac{\alpha_n^*\alpha_{n-1}}{n!(n-1)!} \sum_{k=0}^{n-1}\binom{n-1}{k} \langle0| (\hat B)^n \hat c_1^\dagger (\hat B_{L_1}^\dagger )^k \hat c_{2m}^\dagger(\hat B_{L_2}^\dagger)^{n-1-k}|0\rangle,
\end{align}
where in the second line we have expanded the binomial and commuted $\hat c_{2m}^\dagger$ past all $2k$ creation operators in $(\hat B^\dagger)^k$.

As we argued above, the only nonzero terms in this expression must have contiguous strings of occupied sites connecting site $1$ and site $2m$.
There are two ways to do this: spanning the interval $[1,2m]$ or spanning the interval $[2m,1]$, where the latter wraps around the ring to include site $L$.
We may decompose the correlation function as a sum of these two cases (since the maximally-filled state does not appear in the solution, there is no double-counting of terms).
In doing so, we find that the former requires exactly $L_1/2$ powers of $\hat B_{L_1}^\dagger$, which further reduces to $(\hat B_{L_1}^\dagger)^{L_1/2} = (L_1/2)!\prod_{j=2}^{2m-1}\hat c_j^\dagger$, and the latter requires exactly $L_2/2$ powers of $\hat B_{L_2}^\dagger$, which similarly reduces to $(\hat B_{L_2}^\dagger)^{L_2/2} = (L_2/2)!\prod_{j=2m+1}^{L}\hat c_j^\dagger$.
Here the factorials count the arrangements of the $L_j/2$ dimers. 
The sums over $n$ must now start at $n_{\rm min}^{(1)} = L_1 + 1 = m$ and $n_{\rm min}^{(2)} = L_2 + 1 = L/2 + 1 - m$, respectively ($L_j/2+1$ dimers are needed to include the sites $1$ and $2m$).
Thus, the anomalous correlation is
\begin{align}
    \langle\hat c_1^\dagger \hat c_{2m}^\dagger\rangle 
    & =  \sum_{n=m}^{L/2-1} \frac{\alpha_{n}^{*}\alpha_{n-1}}{n!(n-m)!} \langle0| (\hat B_{L_2} + \hat B_{\rm string})^n (\hat{B}_{L_{2}}^{\dagger})^{n-m} \prod_{j=1}^{2m} (\hat{c}_{j}^{\dagger}) |0\rangle
    \\
    & -  \sum_{n=L/2 - (m-1)}^{L/2-1}\frac{\alpha_{n}^{*}\alpha_{n-1}}{n!(n-L/2+m-1)!} \langle0| (\hat B_{L_1} + \hat B_{\rm string} )^n (\hat{B}_{L_{1}}^{\dagger})^{n-L/2+m-1}\prod_{j=2m}^{L+1}(\hat{c}_{j}^{\dagger})|0\rangle,
    \nonumber
\end{align}
where in each sum we have decomposed $\hat B = \hat B_{L_j} + \hat B_{\rm string}$ where $\hat B_{\rm string}$ includes all sites spanning the contiguous string of occupied sites \emph{including} $1$ and $2m$.
Notice also that in the second sum we have moved $\hat c_1^\dagger \equiv \hat c_{L+1}^\dagger$ to the right to maintain the proper ordering of the string spanning $[2m, 1]$; this picks up a sign in the process as $\hat c_1^\dagger$ commutes past an odd number of creation operators.

The final step is to expand the binomial of $(\hat B)^n$ and to note that the $\hat B_{\rm string}$ term must comprise exactly $m$ powers in the first sum and $L/2+1-m$ powers in the second sum.
These can be reduced to products of fermion annihilation operators similarly to the creation strings.
What remains is a sum of configurations of at most $L_j/2-1$ dimers on the length $L_j$ OBC chains, as the contiguous strings have unit norm (e.g., $\langle0|\hat c_{2m}\cdots \hat c_1\hat c_1^\dagger\cdots\hat c_{2m}^\dagger|0\rangle = 1$).
\begin{align}
    \langle\hat c_1^\dagger \hat c_{2m}^\dagger\rangle 
    &=  \sum_{n=m}^{L/2-1}\frac{\alpha_{n}^{*}\alpha_{n-1}}{[(n-m)!]^{2}}\langle0|(\hat{B}_{L_{2}})^{n-m}(\hat{B}_{L_{2}}^{\dagger})^{n-m}\prod_{j=2m}^{1}(\hat{c}_{j})\prod_{j=1}^{2m}(\hat{c}_{j}^{\dagger})|0\rangle
    \\
    &-  \sum_{n=L/2 - (m-1)}^{L/2-1}\frac{\alpha_{n}^{*}\alpha_{n-1}}{[(n-L/2+m-1)!]^{2}}\langle0|(\hat{B}_{L_{1}})^{n-L/2+m-1}(\hat{B}_{L_{1}}^{\dagger})^{n-L/2+m-1}\prod_{j=L+1}^{2m}(\hat{c}_{j})\prod_{j=2m}^{L+1}(\hat{c}_{j}^{\dagger})|0\rangle
    \nonumber
    \\
    &=  \sum_{n=m}^{L/2-1}\frac{\alpha_{n}^{*}\alpha_{n-1}}{[(n-m)!]^{2}} \Big|\Big|(\hat{B}_{L_{2}}^{\dagger})^{n-m}|0\rangle\Big|\Big|^2
    -  \sum_{n=L/2 - (m-1)}^{L/2-1}\frac{\alpha_{n}^{*}\alpha_{n-1}}{[(n-L/2+m-1)!]^{2}} \Big|\Big| (\hat{B}_{L_{1}}^{\dagger})^{n-L/2+m-1}|0\rangle \Big|\Big|^2
\end{align}
Computing these state norms arrive at an expression for the anomalous correlation functions comprising two finite sums over the state vector coefficients $\alpha_{n}$:
\begin{align}
    \langle\hat c_1^\dagger \hat c_{2m}^\dagger\rangle
    &=  \sum_{n=m}^{L/2-1}\alpha_{n}^{*}\alpha_{n-1}\binom{L-n-m}{n-m}
    -  \sum_{n=L/2-(m-1)}^{L/2-1}\alpha_{n}^{*}\alpha_{n-1}\binom{L/2-n+m-1}{n-L/2+m-1}.
    \label{app-eq:anom-corr-expr}
\end{align}
\end{widetext}
This expression is used to compute the anomalous correlations $\langle\hat c_j^\dagger\hat c_{j+1}^\dagger\rangle$ in Fig.~\ref{fig:phase-diagram}.
The same basic arguments used here can be used to evaluate anomalous correlations in OBC chains and in odd-length PBC chains, but the exact partitions of the chains differ. For OBC chains, the correlations are not translationally invariant, further complicating the calculation.

We make a few remarks about Eq.~\eqref{app-eq:anom-corr-expr}.
First note that for the extremal values of $m = L/2$ or $m = 1$, the first sum or the second sum has inconsistent bounds, respectively; these should evaluate to zero in the respective instances.
This reflects the fact that the maximally-filled state does not appear in $|\psi\cqa\rangle$.
Moreover, for these extremal values, we find the respective correlations evaluate to $\langle\hat c_1^\dagger\hat c_2^\dagger\rangle = L^{-1}\langle\hat B^\dagger\rangle$ and $\langle\hat c_1^\dagger\hat c_L^\dagger\rangle = -L^{-1}\langle\hat B^\dagger\rangle$ (cf.~Eq.~\eqref{app-eq:pair-corr-expr} evaluated for PCB).
This is expected as $\langle\hat B^\dagger\rangle = \sum_{j=1}^L\langle\hat c_j^\dagger\hat c_{j+1}^\dagger\rangle$ is the sum of all $L$ nearest-neighbor anomalous correlations, and translational invariance guarantees that they are uniform.
Finally, note that the expected symmetry $\langle\hat c_j^\dagger\hat c_i^\dagger\rangle = - \langle\hat c_i^\dagger\hat c_j^\dagger\rangle$ does hold.
Note that replacing $m\mapsto L/2 + 1 - m$ is equivalent to swapping sites, up to an irrelevant translation.
Making this replacement in Eq.~\eqref{app-eq:anom-corr-expr}, we find that the two sums interchange, hence the antisymmetry is apparent.
This implies, for example that in a length $L = 4n+2$ PBC chain (e.g., a length $6$ or length $10$ chain), the antipodal anomalous correlations exactly cancel: $\langle\hat c_1^\dagger\hat c_{L/2+1}^\dagger\rangle = 0$; they are the only anomalous correlations that can be nonzero but are exactly zero.


\subsection{Normal correlations in even-length PBC chains}

The reasoning used to compute anomalous correlations can be adapted to compute normal correlations 
\begin{align}
    \langle \hat c_{i}^\dagger\hat c_{j}\rangle \equiv \langle\psi\cqa|\hat c_i^\dagger\hat c_j|\psi\cqa\rangle.
\end{align}
In OBC chains and even-length PBC chains, the only nonzero normal correlations are of the form $\langle\hat c_j^\dagger\hat c_{j+2m}\rangle$ using similar arguments as above regarding contiguous strings of occupied sites.
Note that in these chains, each pair of sites can have either a nonzero anomalous correlation or a nonzero normal correlation, but never both. 
However, as with the anomalous correlations, these constraints do not apply to odd-length PBC chains.

As with the anomalous correlations, translational invariance guarantees $\langle\hat c_j^\dagger\hat c_{j+2m}\rangle = \langle\hat c_1^\dagger\hat c_{2m+1}\rangle$.
Moreover, one can show that the normal correlations are real and symmetric $\langle\hat c_i^\dagger\hat c_j\rangle = \langle\hat c_{j}^\dagger\hat c_i\rangle$.
The calculation of the normal correlations follows similar reasoning to that of the anomalous correlations above, yielding
\begin{widetext}
\begin{align}
    \langle\hat c_1^\dagger\hat c_{2m+1}\rangle = \delta_{m,0} - \sum_{n=m}^{L/2-1} |\alpha_n|^2 \binom{L-n-m-1}{n-m} - \sum_{n=L/2-m}^{L/2-1} |\alpha_n|^2 \binom{L/2-n+m-1}{n-L/2+m}.
\end{align}
\end{widetext}
This expression is used to compute the normal correlations $\langle\hat c_j^\dagger\hat c_{j+2}\rangle$ in Fig.~\ref{fig:phase-diagram}.
The possible values of $m$ range from $m=0$ to $m=L/2 \equiv 0$, which coincide and compute the occupation number.
At these two extremal values, only one sum has consistent bounds, and the other is evaluated as zero. 
One readily verifies that this expression correctly computes the average density $\langle \hat c_1^\dagger\hat c_1 \rangle = L^{-1}\langle \hat N \rangle$.
The symmetry $\langle\hat c_i^\dagger\hat c_j\rangle = \langle\hat c_{j}^\dagger\hat c_i\rangle$ is confirmed by noting that swapping sites $1$ and $2m+1$ is equivalent to the replacement $m\mapsto L/2-m$, up to an irrelevant translation.
Under this replacement, the sums interchange, leaving the expression invariant.


\section{First order phase transition in the thermodynamic limit}
\label{app:phase-transition}

Here we discuss the details of taking the $L\to\infty$ limit of $|\psi\cqa\rangle$ (cf.~Eq.~\eqref{eq:cqa-wavefunc}) and derive the effective free energy $Q(\rho)$ (cf.~Eq.~\eqref{eq:effective-free-energy}) in the weak dissipation limit $\kappa\to0$ and for finite dissipation.
We begin by writing the CQA wavefunction in terms of normalized states
\begin{align}
    |\psi\cqa\rangle &= \sum_{n=0}^{L/2-1} \beta_n|n\rangle,\quad
    |n\rangle \equiv \frac{(\hat B^\dagger)^n}{n!\sqrt{\mathcal{N}(L,n)}}|0\rangle,
\end{align}
(cf.~Eqs.~\eqref{eq:cqa-beta} and \eqref{eq:n-states}).
The states $|n\rangle$ are orthonormal states ($\langle m|n\rangle = \delta_{mn}$) of $n$ fermion pairs distributed according to $\hat B^\dagger$ (i.e., nearest neighbor pairs uniformly delocalized across the lattice for the 1D $p$-wave pairing model).
The \emph{normalized} state vector coefficients $\beta_n=\alpha_n\sqrt{\mathcal{N}(L,n)/\mathcal{N}\cqa}$ are explicitly given by
\begin{align}
    \beta_n = \frac{1}{\sqrt{\mathcal{N}\cqa}} \frac{(L\Delta)^n}{\prod_{m=1}^n [L\tilde{\mu} - m]} \Big [ \frac{L}{n}\binom{L-n-1}{n-1} \Big]^{1/2},
\end{align}
where we have taken $E_C=1$ (cf.~Eq.~\eqref{eq:alpha-n-EC1}) and factored out a $1/L$ from each term in the product.
The product is now the form of a falling factorial and, thus, can be written in terms of the ratio of $\Gamma$ functions,
\begin{align}
    \prod_{m=1}^n [L\tilde{\mu} - m] = \frac{\Gamma(L\tilde{\mu})}{\Gamma(L\tilde{\mu} - n)}.
\end{align}
For any $\kappa>0$, the arguments have nonzero imaginary parts, so no poles of $\Gamma(z)$ are encountered in this expression.

Now, we introduced the density coordinate
\begin{align}
    \rho \equiv \frac{2n}{L},\quad 0 < \rho < 1,
\end{align}
which becomes a continuous parameter in the large $L$ limit.
Making this change of variable in $\beta_n\to\beta(\rho)$, then using Stirling's approximation for $\Gamma(z)$ and the binomial coefficients, we find
\begin{widetext}
\begin{align}
    \beta(\rho) = \mathcal{N}\cqa^{-1/2}\frac{2}{\rho} \sqrt{\frac{L\rho(1-\rho)}{2\pi(2-\rho)(1-\frac{\rho}{2\tilde\mu})}}  \left[ \left( \frac{e\Delta}{\tilde\mu(1- \frac{\rho}{2\tilde\mu})} \right)^{\rho/2} \left(1 - \frac{\rho}{2\tilde\mu}\right)^{\tilde\mu} \left( 1 - \frac{\rho}{2} \right)^{1-\rho/2} \left(\frac{\rho}{2}\right)^{-\rho/2} \left( 1-\rho\right)^{-(1-\rho)}  \right]^L
    \label{app-eq:beta-rho}
\end{align}
\end{widetext}
where the normalization is now $\mathcal{N}\cqa = \int_0^1d\rho|\beta(\rho)|^2$.


\subsection{Effective free energy}

The density probability distribution $P(\rho) = |\beta(\rho)|^2$, for $\beta(\rho)$ given by Eq.~\eqref{app-eq:beta-rho}, is system-size-dependent.
However, in the limit $L\to\infty$, its functional form is dominated by the bracketed factor raised to the power of $L$.
Thus, we can define the effective free energy
\begin{align}
    Q(\rho) = -\lim_{L\to\infty} L^{-1} \ln P(\rho),
\end{align}
which has a well-defined limit:
\begin{align}
    Q&(\rho) =
    \label{app-eq:effective-free-energy-full}
    \\
    & -(1 + \ln \Delta)\rho - \left( \mu - \frac{\rho}{2} \right) \ln \left[ \left(\mu - \frac{\rho}{2} \right)^2 +\frac{\kappa^2}{4} \right] 
    \nonumber
    \\
    &+ \mu \ln \left( \mu^2 + \frac{\kappa^2}{4} \right) + \kappa \arctan\left( \frac{\rho\kappa/2}{2|\tilde\mu|^2 - \mu\rho} \right)
    \nonumber
    \\
    &- \left(1 - \frac{\rho}{2}\right) \ln \left(1 - \frac{\rho}{2}\right) + (1-\rho) \ln (1-\rho) + \frac{\rho}{2} \ln \frac{\rho}{2}.
    \nonumber
\end{align}
Here the arctangent expresses $\arg(1-\rho/2\tilde\mu)$; the individual signs of the numerator and denominator determine the evaluation quadrant.
The probability distribution for finite $L\gg 1$ is well-approximated by $P(\rho) \approx e^{-LQ(\rho)}/\int_0^1 d\rho e^{-LQ(\rho)}$, and is given exactly by this expression in the limit.
For the remainder of this section, we take the weak dissipation limit $\kappa\to 0$, for which the effective free energy reduces to
\begin{align}
    &\lim_{\kappa\to 0^+} Q(\rho)= 
    \label{app-eq:effective-free-energy-weak-diss}
    \\
    &\quad - (1 + \ln \Delta)\rho - 2\left( \mu - \frac{\rho}{2} \right) \ln \left| \mu - \frac{\rho}{2}\right| + 2\mu \ln |\mu|
    \nonumber
    \\
    &\quad - \left(1 - \frac{\rho}{2}\right) \ln \left(1 - \frac{\rho}{2}\right) + (1-\rho) \ln (1-\rho) + \frac{\rho}{2} \ln \frac{\rho}{2}.
    \nonumber
\end{align}
We discuss finite dissipation in the next section.

Through examination of $Q(\rho)$ and its derivatives, we find that for generic parameters $\Delta>0$ and $\mu\in\mathbb{R}$, the effective free energy has a single minimum $\min_{\rho\in [0,1]} Q(\rho) = Q(\rho_{\rm min})$, and in the thermodynamic limit, the probability distribution evaluates to $P(\rho) = \delta(\rho - \rho_{\rm min})$.
This can be rigorously shown by expanding $Q(\rho-\rho_{\rm min}) = Q(\rho_{\rm min}) + \frac{1}{2}Q^{\prime\prime}(\rho_{\rm min})(\rho - \rho_{\rm min})^2$. 
The support of $Q(\rho-\rho_{\rm min})$ lies in a finite region $\sim [{\rho_{\rm min} - L^{-1}}, {\rho_{\rm min} + L^{-1}}]$ as it vanishes exponentially outside this region, so we may take the limits of the normalization integral to $\pm\infty$ with exponentially small error $\mathcal{O}(e^{-L})$.
Thus,
\begin{align}
    P(\rho) &= \lim_{\epsilon\to0} \frac{\exp\big[-\frac{1}{2\epsilon^2} (\rho - \rho_{\rm min})^2 \big]}{\int_{-\infty}^{\infty} d\rho \exp\big[-\frac{1}{2\epsilon^2} (\rho - \rho_{\rm min})^2 \big]}
    \label{app-eq:P-delta-func}
    \\
    &= \lim_{\epsilon\to0} \frac{1}{\epsilon\sqrt{2\pi}}e^{-\frac{1}{2\epsilon^2}(\rho - \rho_{\rm min})^2} \nonumber
\end{align}
where $\epsilon =  1/\sqrt{LQ^{\prime\prime}(\rho_{\rm min})}$, and we recognize the limit in the second line as the Gaussian definition of $\delta(\rho-\rho_{\rm min})$.

In the specific region $0<\mu<0.5$, and for sufficiently small $\Delta>0$ (here, the sufficiently small region always includes $\Delta_{\rm crit}(\mu)$ for the given $\mu$), the effective free energy has two local minima for $\rho\in[0,1]$.
The minima always lie on opposite sides of $\rho=2\mu$, and we define them by
\begin{align}
    Q(\rho_{\rm low}) &= \min_{\rho\in[0,2\mu]}Q(\rho),
    \\
    Q(\rho_{\rm high}) &= \min_{\rho\in[2\mu,1]}Q(\rho).
\end{align}
The difference in potential of the two minima, $\delta Q_{\rm min} \equiv Q(\rho_{\rm high}) - Q(\rho_{\rm low})$, is a monotonic function of both $\mu$ and $\Delta$, decreasing with $\mu$ and increasing with $\Delta$.
For any finite potential difference $|\delta Q_{\rm min}|>0$, the lower potential minima will dominate as $L\to\infty$, as for any fixed $\epsilon>0$, $\lim_{L\to\infty}L\epsilon = \infty$ and the argument above Eq.~\eqref{app-eq:P-delta-func} applies.
Thus, 
\begin{align}
    P(\rho) = 
    \begin{cases}
        \delta(\rho - \rho_{\rm low}) & \delta Q_{\rm min}>0
        \\
        \delta(\rho - \rho_{\rm high}) & \delta Q_{\rm min}<0
    \end{cases},
\end{align}
so there will be a jump discontinuity in the particle density as the sign of the potential difference changes.
Because $\delta Q_{\rm min}$ is monotonic in both $\mu$ and $\Delta$, this discontinuity will occur if the system crosses the critical line $\Delta_{\rm crit}(\mu)$ in any direction in the $\mu$--$\Delta$ plane.


\subsection{Finite dissipation}

When we consider finite dissipation, we must use the full effective free energy Eq.~\eqref{app-eq:effective-free-energy-full} instead of Eq.~\eqref{app-eq:effective-free-energy-weak-diss}.
The qualitative features of the phase transition, including the jump discontinuity of particle density across the critical line, remain intact.
The thermodynamic limit probability distribution remains $P(\rho) = \delta(\rho-\rho_{\rm min})$, and near the critical line, there remains two local minima of $Q(\rho)$ at $\rho_{\rm low}<2\mu$ and $\rho_{\rm high}>2\mu$.
The jump discontinuity in particle density still occurs when $\delta Q_{\rm min} \equiv Q(\rho_{\rm high}) - Q(\rho_{\rm low})$ changes sign. 

\begin{figure}[t!]
	\centering
	\includegraphics[]{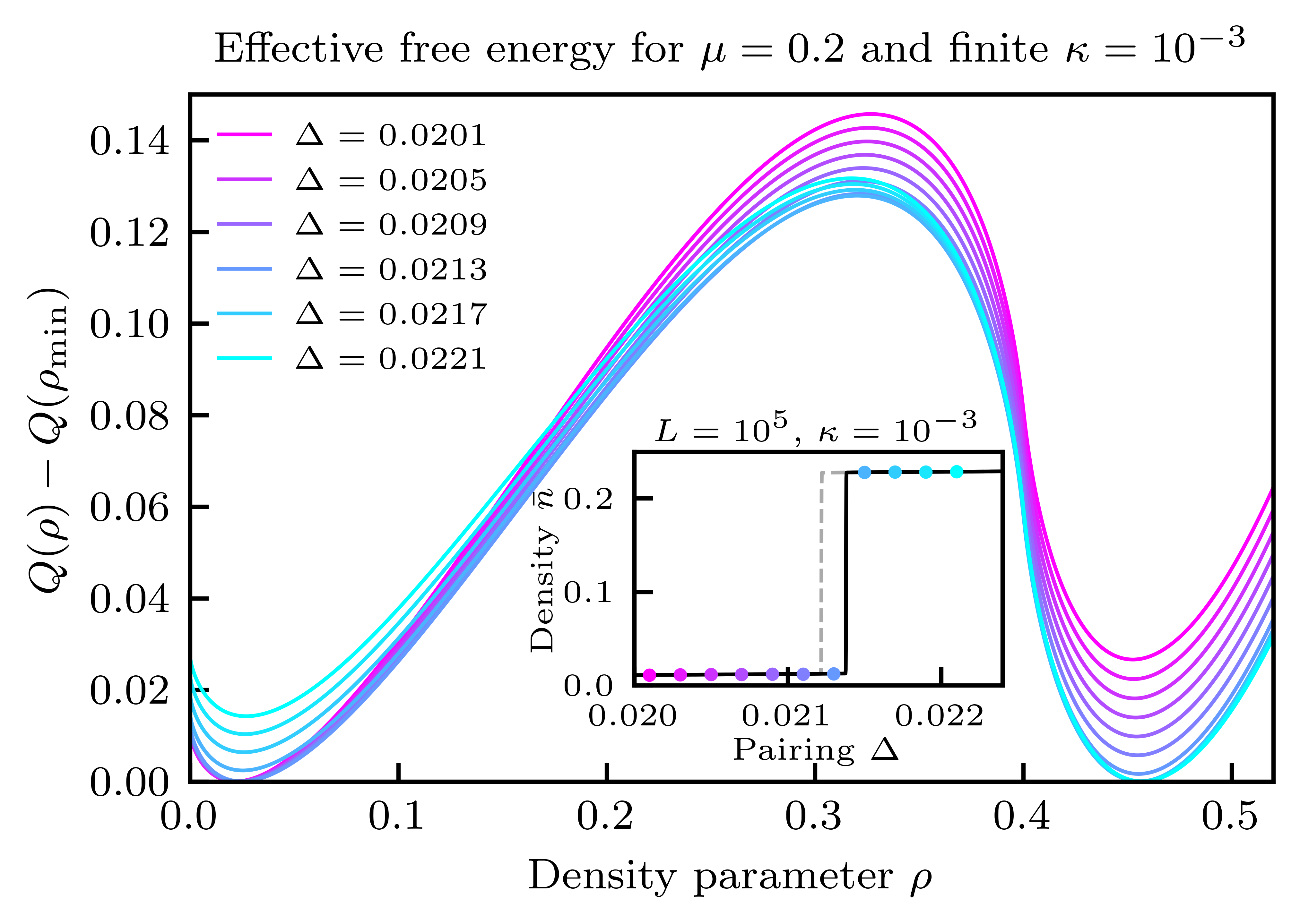}
	\caption{
	\textbf{Effective free energy and phase transition for finite dissipation.}
    The effective free energy $Q(\rho)$ given by Eq.~\eqref{app-eq:effective-free-energy-full} is shown at chemical potential $\mu = 0.2$ and at finite $\kappa = 10^{-3}$ for pairing potentials ranging from $\Delta = 0.0201$ to $\Delta = 0.0223$ (each successive curve steps by $0.002$).
    For each $\Delta$, the minimum $Q(\rho_{\rm min})$ is subtracted off so that each curve touches zero at exactly one point.
    The phase transition occurs at $\Delta_{\rm crit}\approx 0.02138$ (in the weak dissipation limit $\Delta_{\rm crit}\approx 0.02122$, cf.~Fig.~\ref{fig:phase-transition}).
    \textbf{Inset:}
    The exact physical particle density for a $L=10^5$ system is plotted as a function of $\Delta$ for $\mu = 0.2$ and $\kappa = 10^{-3}$ around the phase transition. Overlaid as individual points are the densities $\bar n = \rho_{\rm min}/2$ predicted from $\min Q(\rho)$ at their respective $\Delta$. The dashed grey curve is the particle density for $\kappa = 10^{-8}$, as shown in Fig.~\ref{fig:phase-transition}.
    }
	\label{app-fig:phase-transition-finite-kappa}
 \end{figure}

There are two principal quantitative effects of finite dissipation: the phase boundary $\Delta_{\rm crit}(\mu)$ is weakly renormalized by $\kappa$, and the phase boundary terminal point, which $\mu=0.5$ in the weak dissipation limit, moves to a smaller value $\mu<0.5$.
In Fig.~\ref{app-fig:phase-transition-finite-kappa}, we demonstrate the phase boundary renormalization effect.
The parameters for this figure are identical to that of Fig.~\ref{fig:phase-transition} of the main text.


\section{Mean-field analysis}
\label{app:mean-field}

The master equation of the 1D chain with $p$-wave pairing in momentum space (annihilation operators $\hat c_k = L^{-1/2}\sum_j e^{ijk}\hat c_j$) is
\begin{align}
    \partial_t &\hat\rho = -i[\hat H,\hat\rho] + \kappa \sum_k \mathcal{D}[\hat c_k]\hat\rho,
    \label{app-eq:momentum-qme}
    \\
    \hat H &= -\mu \sum_k \hat n_k + \frac{E_C}{2L}\Big(\sum_{k}\hat n_k\Big)^2 + \sum_k \Delta_k\big(\hat c_k^\dagger\hat c_{-k}^\dagger + {\rm H.c.}),
    \label{app-eq:momentum-H}
\end{align}
where $\Delta_k = \Delta\sin k$ is the amplitude for the pairing of momentum modes $k$ and $-k$. 
Note that we have gauged away a phase of $i$ on $\Delta_k$ so that it is real.

To treat the interaction $\propto E_C$ at the mean-field level, we assume that all bilinear operators have self-consistently determined expectation values: $\hat c_k^\dagger\hat c_{k\prime} \to \hat c_k^\dagger\hat c_{k\prime} + \langle \hat c_k^\dagger\hat c_{k\prime}\rangle$ and $\hat c_k \hat c_{k^\prime} \to \hat c_k \hat c_{k^\prime} + \langle \hat c_k \hat c_{k^\prime} \rangle$.
Therefore, we consider all possible ``contractions'' of $\hat H_{\rm int} = (E_C/2L)(\sum_k\hat n_k)^2$ into a quadratic term multiplied by a quadratic expectation value.
Neglecting the quartic fluctuations, the mean-field interaction is
\begin{align}
    \hat H_{\rm int}^{\rm (mf)} &= E_C\bar n\sum_{q}\hat{n}_{q} + \frac{E_{C}}{L}\sum_{q\neq q^{\prime}}\langle\hat{c}_{q}^{\dagger}\hat{c}_{q^{\prime}}\rangle\hat{c}_{q^{\prime}}^{\dagger}\hat{c}_{q} 
    \\
    &+ \frac{E_{C}}{2L}\sum_{qq^{\prime}}\left[\langle\hat{c}_{q^{\prime}}^{\dagger}\hat{c}_{q}^{\dagger}\rangle\hat{c}_{q}\hat{c}_{q^{\prime}}+\mathrm{h.c.}\right],
    \nonumber
\end{align}
where $\bar{n} = L^{-1}\sum_k \langle\hat n_k\rangle$ is the mean density.
Thus, our mean-field master equation is given by Eq.~\eqref{app-eq:momentum-qme} with the interaction term $\propto E_C$ replaced by $\hat H_{\rm int}^{\rm (mf)}$.

In principle, we can self-consistently compute all quadratic mean-field quantities; however, we focus on the mean density $\bar{n} = L^{-1}\sum_k\langle\hat n_k\rangle$.
We find that the equations of motion for the $\langle\hat n_k\rangle$ depend only on the anomalous correlations between $\pm k$ modes:
\begin{align}
    \partial_{t}\langle\hat{n}_{k}\rangle
    &=-\kappa\langle\hat{n}_{k}\rangle-2i\Delta\left(\langle\hat{c}_{k}^{\dagger}\hat{c}_{-k}^{\dagger}\rangle-\langle\hat{c}_{-k}\hat{c}_{k}\rangle\right).
    \label{app-eq:nk-eom}
\end{align}
All contributions to the equations of motion from the normal correlation $\langle\hat c_k^\dagger\hat c_q\rangle$ exactly cancel out, as do those from the anomalous correlations between any mode pairs other than $\pm k$.
A similar calculation shows a similar result for the anomalous correlations: 
\begin{align}
    \partial_{t}\langle\hat{c}_{k}^{\dagger}\hat{c}_{-k}^{\dagger}\rangle&=-2i\Big(\mu-E_C\bar n-\frac{i\kappa}{2}-\frac{E_{C}}{2L}\Big)\langle\hat{c}_{k}^{\dagger}\hat{c}_{-k}^{\dagger}\rangle
    \\
    &-2i\Big(\Delta\sin k-\frac{E_{C}}{2L}\langle\hat{c}_{k}^{\dagger}\hat{c}_{-k}^{\dagger}\rangle\Big)\big(\langle\hat{n}_{k}\rangle+\langle\hat{n}_{-k}\rangle\big)
    \nonumber
    \\
    &+2i\Delta\sin k.
    \nonumber
\end{align}
Again, all contributions from the normal correlations exactly cancel.
Due to the inversion symmetry of the model ($x\to -x$ and $k\to-k$) we have $\langle\hat n_k\rangle = \langle\hat n_{-k}\rangle$, and upon taking the $L\to\infty$ limit, this express further simplifies to
\begin{align}
    \partial_{t}\langle\hat{c}_{k}^{\dagger}\hat{c}_{-k}^{\dagger}\rangle&=-2i\Big(\mu - E_C\bar n - \frac{i\kappa}{2}\Big) \langle\hat{c}_{k}^{\dagger}\hat{c}_{-k}^{\dagger}\rangle
    \label{app-eq:anomk-eom}
    \\
    &\quad -4i \Delta\sin k \Big(\langle\hat{n}_{k}\rangle - \frac{1}{2}\Big).
    \nonumber
\end{align}
Note the only appearance of the mean-field interaction $\propto E_C\bar n$ is in this equation of motion.

Solving for the steady state by setting $\partial_t\langle\hat n_k\rangle = \partial_t \langle\hat c_k^\dagger\hat c_{-k}^\dagger\rangle = 0$ in Eqs.~\eqref{app-eq:nk-eom} and \eqref{app-eq:anomk-eom}, we find
\begin{align}
    \langle\hat{n}_{k}\rangle &= \frac{2\Delta^{2}\sin^{2}k} {\left(\mu - E_C\bar n\right)^{2} + \kappa^{2}/4 + 4\Delta^{2}\sin^{2}k}.
\end{align}
From this, we obtain the self-consistency equation for the mean density $\bar n = L^{-1}\sum_k \langle\hat n_k\rangle$.
In the $L\to\infty$ limit, we replace the sum with an integral, $L^{-1}\sum_{k}\to\int_{-\pi}^{\pi}\frac{dk}{2\pi}$, which we can perform analytically to obtain the self-consistency equation
\begin{align}
    \bar{n}=\frac{1}{2}\left(1-\frac{\sqrt{(E_C\bar n-\mu)^{2}+\kappa^{2}/4}}{\sqrt{(E_C\bar n-\mu)^{2}+\kappa^{2}/4+4\Delta^{2}}}\right).
\end{align}


\section{Mapping to the mean-field transverse-field Ising model}
\label{app:tfim}

Here, we discuss the details and subtleties of mapping the fermion master equation to a transverse-field Ising model. The Anderson pseudospin mapping for spinless fermion is \cite{anderson_RandomPhase_1958}
\begin{align}
    \hat \sigma_k^{-} = \hat c_{-k}\hat c_k,\quad \hat \sigma_k^z = \hat n_k +\hat n_{-k} -1
\end{align}
where the momentum index is restricted to $0\leq k < \pi$ to avoid double counting, and the $k=0$ and $k=\pi$ modes are paired to form the $k=0$ spin.
The mapping of the full interacting Hamiltonian Eq.~\eqref{app-eq:momentum-H} is exact; a few lines of simple algebra yields
\begin{align}
    \hat H_{\rm spin} &= -\mu\sum_{k} \hat \sigma_k^z + \frac{E_C}{2L}\sum_{k,k^\prime}(\hat\sigma_k^z +1) (\hat\sigma_{k^\prime}^z +1) 
    \label{app-eq:spin-H-int}\\
    &\qquad+ \sum_{k}2\Delta_k\hat\sigma_k^x, \nonumber
\end{align}
where $0\leq k < \pi$ is assumed we in all sums and $\Delta_k = \Delta\sin k$ (hence, the $k=0$ pair is undriven).
Likewise, the mean-field Hamiltonian (with mean-field interaction $\hat H_{\rm int}^{\rm (mf)} = \bar{n}E_C\sum_k\hat n_k$ with $\bar{n}=L^{-1}\sum_k\langle\hat n_k\rangle$) exactly maps to the mean-field spin Hamiltonian
\begin{align}
    \hat H_{\rm spin}^{\rm (mf)} &= \Big(-\mu + \frac{E_C}{2}(\bar{M}+1) \Big)\sum_{k} \hat \sigma_k^z + \sum_{k}2\Delta_k\hat\sigma_k^x,
\end{align}
where $\bar{M} = (L/2)^{-1}\sum_k\langle\sigma_k^z\rangle$ is the mean magnetization.
One readily shows that $(\bar{M}+1)/2 = \bar{n}$, hence there is a perfect mapping from the $\bar{n}$-renormalized fermion chemical potential to the $\bar{M}$-renormalized spin longitudinal field.

\subsection{Mean-field expectation value equivalence}

Unlike the Hamiltonians, the single particle fermion dissipation does not have an exact mapping to spins as there is no way to represent $\hat c_k$ in terms of the spin operators.
Nevertheless, we can show that for an initial state $\hat \rho_0$ in the pseudospin subspace, there is a correspondence between equations of motion for the spin operators and the fermion operators in the mean-field theory (and, thus, noninteracting limit $E_C=0$) using the dissipation mapping Eq.~\eqref{eq:dissipation-equivalence}.
That is, given an initial state in the pseudospin subspace, and a fermion operator which exactly maps to a spin operator $\hat{O}_{\rm fermi} \simeq \hat{O}_{\rm spin}$, each subject to time evolution under their respective fermion and spin Lindbladians 
\begin{align}
    \mathcal{L}_{\rm fermi} &= -i[\hat H_{\rm fermi}^{\rm (mf)},\cdot] + \kappa\sum_{k>0}\Big(\mathcal{D}[\hat c_k] + \mathcal{D}[\hat c_{-k}]\Big),
    \label{app-eq:L-fermi}
    \\
    \mathcal{L}_{\rm spin} &= -i[\hat H_{\rm spin}^{\rm (mf)},\cdot] + \kappa\sum_{k>0}\Big(\mathcal{D}[\hat \sigma^-_k] + \frac{1}{4}\mathcal{D}[\hat c^z_{k}]\Big),
    \label{app-eq:L-spin}
\end{align}
where $\hat H_{\rm fermi}^{\rm (mf)}$ is given by mean-field approximation to Eq.~\eqref{app-eq:momentum-H}, the equations of motion for the operator expectation values are equivalent:
\begin{align}
    \langle\hat{O}_{\rm fermi}(t)\rangle = \langle\hat{O}_{\rm spin}(t)\rangle,
    \label{app-eq:operator-equivalence}
\end{align}
where $\langle\hat{O}_{\rm s}(t)\rangle = {\rm Tr} [\hat{O}_{\rm s} \hat\rho_{\rm s}(t)] = {\rm Tr} [\hat{O}_{\rm s} (e^{t\mathcal{L}_{\rm s}}\hat\rho_{0,{\rm s}})]$ for the respective system.
Here an initial state in the pseudospin subspace is simply any spin density matrix $\hat\rho_{0,{\rm spin}}$ which has been mapped back to fermions. It comprises only paired fermion populations (e.g., $\hat c_{k}^\dagger\hat c_{-k}^\dagger|0\rangle\langle 0|\hat c_{-k} \hat c_{k}$) and coherences (e.g., $\hat c_{k}^\dagger\hat c_{-k}^\dagger|0\rangle\langle 0| + {\rm H.c.}$).

Given $\hat \rho_0$ in the pseudospin subspace, the initial expectation values coincide $\langle \hat O_{\rm fermi}(0)\rangle = \langle \hat O_{\rm fermi}(0)\rangle$. It is therefore sufficient to show that the equations of motion are identical: $\partial_t \langle\hat O_{\rm fermi}\rangle = \partial_t\langle \hat O_{\rm spin}\rangle$, where $\partial_t\langle\hat O_{\rm s}\rangle = {\rm Tr}[\hat O_{\rm s}(\mathcal{L}_{\rm s}\hat\rho_{\rm s})]$.

The expectation values of the linear spin operators $\hat \sigma_k^-$ and $\hat \sigma_k^z$ obey the equations of motion
\begin{align}
    \partial_t \langle\hat\sigma_k^-\rangle &= \big(2i\mu - i E_C(\bar{M}+1) - \kappa \big)\langle\hat\sigma_k^-\rangle + 2i\Delta_k\langle\hat\sigma_k^z\rangle,
    \nonumber
    \\
    \partial_t \langle\hat\sigma_k^z\rangle &= - \kappa\langle\hat\sigma_k^z\rangle - \kappa + 4i\Delta_k(\langle\hat\sigma_k^-\rangle -\langle\hat\sigma_k^+\rangle ).
\end{align}
Here, $\langle\hat\sigma_k^-\rangle$ decays with rate $\kappa/2$ due to the loss $\mathcal{D}[\hat\sigma_k^-]$ and with rate $\kappa/2$ due to the dephasing $\mathcal{D}[\hat\sigma_k^z]$, and $\langle\hat\sigma_k^z\rangle$ decays with rate $\kappa$ entirely due to the loss.
For the fermions, inversion symmetry implies that $\partial_t\langle\hat n_k\rangle = \partial_t\langle\hat n_{-k}\rangle$, so it is sufficient to consider the equations of motion for $\langle\hat c_{-k}\hat c_k\rangle$ and $\langle\hat n_k \rangle$,
\begin{align}
    \partial_t \langle \hat c_{-k}\hat c_k\rangle &= \big( 2i(\mu - E_C \bar{n}) -\kappa \big) \langle \hat c_{-k}\hat c_k\rangle \nonumber \\
    &\qquad\qquad- 2i\Delta_k \big(\langle \hat n_k \rangle + \langle \hat n_{-k}\rangle -1\big),
    \nonumber
    \\
    \partial_t \langle \hat n_k\rangle &= - \kappa\langle \hat n_k \rangle + 4i\Delta_k \big(\langle\hat c_{-k}\hat c_k\rangle - \langle\hat c_{k}^\dagger\hat c_{-k}^\dagger\rangle \big).
\end{align}
Here, $\langle \hat c_{-k}\hat c_k\rangle$ decays with rate $\kappa/2$ due to each $\pm k$ mode loss for a total decay rate of $\kappa$, and $\langle\hat n_k\rangle$ decays with rate $\kappa$ due to the loss on that mode.

\begin{figure}[t!]
	\centering
	\includegraphics[]{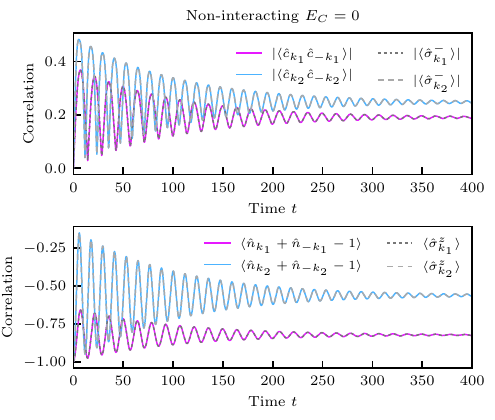}
	\caption{
	\textbf{TFIM mapping of the noninteracting fermion model.}
    The dynamics of fermion operators evolving under Eq.~\eqref{app-eq:L-fermi} are compared with the dynamics of their spin equivalents evolving under Eq.~\eqref{app-eq:L-spin} in the non-interacting limit $E_C=0$. Here we numerically simulate a $L=10$ fermion system ($N=5$ spin system) with parameters $\kappa=0.01$, $\mu = 0.2$, and $\Delta=0.3$ (in arbitrary units). In the top panel we compare $\langle\hat c_k \hat c_{-k}\rangle$ to their equivalents $\langle\hat\sigma_k^-\rangle$, and in the lower panel we compare $(\langle\hat n_k\rangle + \langle\hat n_{-k}\rangle -1)$ to $\langle\hat\sigma_k^z\rangle$. For both panels, $k_1 = \pi/10$ and $k_2 = \pi/5$. As we expect, the equivalence Eq.~\eqref{app-eq:operator-equivalence} holds.
    Although not shown here, we find the equivalence also holds for higher moments (e.g., $\langle\hat \sigma_{k_1}^z\sigma_{k_2}^z\rangle$).
    }
	\label{app-fig:tfim-free}
 \end{figure}

Using the mean-field density relation $\bar{n} = (\bar{M}+1)/2$ and the pseudospin mapping $\langle\hat\sigma_k^z\rangle = \langle\hat n_k\rangle + \langle\hat n_{-}\rangle - 1$, we find that the mean-field equations of motion for the spin and fermion models are identical.
Moreover, because the mean-field theory is quadratic, the equivalence holds for all higher moments by Wick's theorem.
We verify the result numerically for a $L=10$ noninteracting fermion system ($N=5$ spin system) in Fig.~\ref{app-fig:tfim-free}.

\subsection{Breakdown of the mapping with interactions}

To illustrate how the mapping breaks down for the interacting system, it is sufficient to consider a single momentum pair $\pm k$ initialized in an arbitrary pseudospin state
\begin{align}
    \hat\rho_0 &= (1-p)|0\rangle\langle0| + p\hat c_{k}^\dagger\hat c_{-k}^\dagger|0\rangle\langle0|\hat c_{-k}\hat c_k \\
    &+ \big( \beta\hat c_{k}^\dagger\hat c_{-k}^\dagger|0\rangle\langle0| + {\rm H.c.} \big),\nonumber
\end{align}
where $0\leq p\leq 1$ is the pair population and $|\beta|\leq \sqrt{p(1-p)}$ is the pair coherence.
The equivalent spin state is
\begin{align}
    \hat\rho_0 = (1-p)|0\rangle\langle0| + p\hat \sigma^+_{k}|0\rangle\langle0|\hat \sigma_{k}^- + \left( \beta\hat \sigma_{k}^+|0\rangle\langle0| + {\rm H.c.} \right).
\end{align}
For the sake of clarity, we consider the evolution of these states under the dissipation and the nonlinearity alone.
Crucially, the fermion system has a true nonlinearity but the spin system does not:
\begin{align}
    \hat H_{\rm fermi} &= \frac{E_C}{4}(\hat n_k +\hat n_{-k} + \hat n_k\hat n_{-k}),
    \\
    \hat H_{\rm spin} &= \frac{E_C}{2}(\hat \sigma_k^z + 1).
\end{align}
For the initial state in the pseudospin subspace, the expectation of these two Hamiltonians coincide,
\begin{align}
    \langle\hat H_{\rm fermi}\rangle_0 = \langle\hat H_{\rm spin}\rangle_0 = E_C \alpha^2.
\end{align}
Moreover, an initial state in the pseudospin subspace remains in the subspace under Hamiltonian evolution so these expectation values would continue to coincide under purely Hamiltonian evolution.

The correspondence breakdown occurs because the dissipation takes the fermion system out of the pseudospin subspace, and the fermion interaction behaves differently from its spin mapping out of the subspace.
We consider the evolution of the fermion and spin states over a short time $\delta t$: $\rho_{\rm s}(\delta t) = \rho_{0,{\rm s}} + \delta t\mathcal{L}_{\rm s}\hat\rho_{0,{\rm s}}$.
In fact, it is enough to consider only the change in the state $\Delta\hat\rho_{\rm s}/\delta t = \mathcal{L}_{\rm s}\hat\rho_{0,{\rm s}}$.
Thus,
\begin{align}
    \Delta\hat\rho_{\rm fermi} &= \big(-iE_{C}-\kappa\big)\beta\hat{c}_{k}^{\dagger}\hat{c}_{-k}^{\dagger}|0\rangle\langle0|
    \\
    &+\big(iE_{C}-\kappa\big)\beta^{*}|0\rangle\langle0|\hat{c}_{-k}\hat{c}_{k} \nonumber
    \\
    &+\kappa p\Big(\hat{c}_{k}^{\dagger}|0\rangle\langle0|\hat{c}_{k}+\hat{c}_{-k}^{\dagger}|0\rangle\langle0|\hat{c}_{-k}\Big) \nonumber
    \\
    &-2\kappa p \hat{c}_{k}^{\dagger}\hat{c}_{-k}^{\dagger}|0\rangle\langle0|\hat{c}_{-k}\hat{c}_{k},\nonumber
    \\
    \Delta \hat\rho_{\rm spin} &= \big(-iE_{C}-\kappa\big)\beta\hat{\sigma}_{k}^{+}|0\rangle\langle0|
    \\
    &+\big(iE_{C}-\kappa\big)\beta^{*}|0\rangle\langle0|\hat{\sigma}_{k}^{-} \nonumber
    \\
    &+\kappa p\Big(|0\rangle\langle0|-\hat{\sigma}_{k}^{+}|0\rangle\langle0|\hat{\sigma}_{k}^{-}\Big). \nonumber
\end{align}
The coherences ($\propto \beta,\beta^*$) evolve identically, but the populations do not.
In the spin system, the excited state population is reduced and the ground state population is increased by the same amount.
In the fermion system, the pair population is reduced by twice the amount and that population is transferred into each of the single fermion modes equally; this is a manifestation of the pair-breaking effect of the single-particle loss.

\begin{figure}[t!]
	\centering
	\includegraphics[]{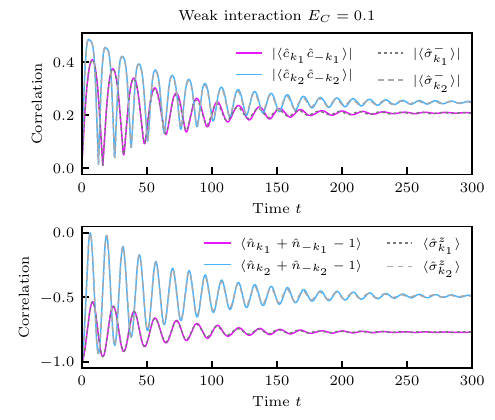}
    \includegraphics[]{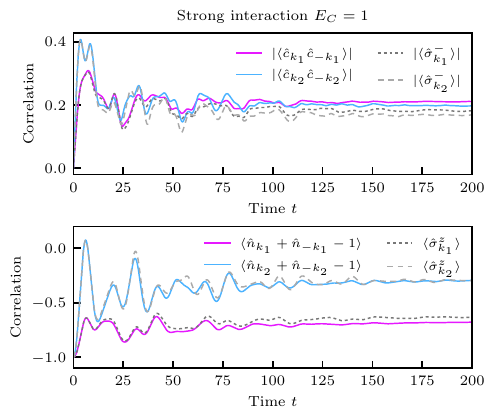}
	\caption{
	\textbf{TFIM mapping of the interacting fermion model.}
    The dynamics of fermion operators evolving under Eq.~\eqref{app-eq:L-fermi}, with the interacting fermion Hamiltonian Eq.~\eqref{app-eq:momentum-H}, are compared with the dynamics of their spin equivalents evolving under Eq.~\eqref{app-eq:L-spin}, with the interacting spin Hamiltonian Eq.~\eqref{app-eq:spin-H-int}. Here we numerically simulate a $L=10$ fermion system ($N=5$ spin system) with parameters $\kappa=0.01$, $\mu = 0.2$, and $\Delta=0.3$ (in arbitrary units). The top two panels are for ``weak'' interactions $E_C=0.1$ and the bottom two panels are for ``strong'' interactions $E_C=1$.
    In the top panel of each interaction regime we compare $\langle\hat c_k \hat c_{-k}\rangle$ to their equivalents $\langle\hat\sigma_k^-\rangle$, and in the lower panel we compare $(\langle\hat n_k\rangle + \langle\hat n_{-k}\rangle -1)$ to $\langle\hat\sigma_k^z\rangle$. For all panels, $k_1 = \pi/10$ and $k_2 = \pi/5$. The parameters used and operators shown are identical to those of Fig.~\ref{app-fig:tfim-free}.
    }
	\label{app-fig:tfim-interacting}
 \end{figure}

We find that the change in linear spin expectation values coincide with their quadratic counterparts: $\Delta\langle\hat\sigma_k^-\rangle = \Delta\langle\hat c_{-k}\hat c_k\rangle = (-iE_C-\kappa)\beta$ and $\Delta\langle\hat\sigma_k^z\rangle = \Delta\langle\hat n_k +\hat n_{-k}-1\rangle = -2\kappa p$.
On the other hand, the difference in how the dissipation alters the populations manifests in different expectation values for the interaction Hamiltonians:
\begin{align}
    \Delta\langle\hat{H}_{\mathrm{fermi}}\rangle&=-\frac{3}{2}\kappa pE_{C},
    \\
    \Delta\langle\hat{H}_{\mathrm{spin}}\rangle&=-\kappa pE_{C}.
\end{align}
This implies that in the next time step, these two equivalent observables will differ evolve differently.
This is entirely caused by the way the fermionic charging energy is sensitive to the system leaving the pseudospin subspace.
Alternatively, one readily shows that $[\hat H_{\rm fermi},\Delta\hat\rho_{\rm fermi}] \neq [\hat H_{\rm spin},\Delta\hat\rho_{\rm spin}]$; thus, in the Schr{\"o}dinger picture, the two systems' density matrices evolve differently.
Finally, we show the correspondence breakdown numerically in Fig.~\ref{app-fig:tfim-interacting}.

\section{Generating function solution to the dimer counting problem}
\label{app:gen-func}

In this appendix, we give a review of the exact solution of the dimer covering problem discussed in Ref.~\cite{fisher_Association_1960}. For an open chain of length $L$ we seek the generating function
\begin{align}
    F_{L}(c_{0},c_{1}) = \sum_{n}N(L,n)c_{0}^{L-2n}c_{1}^{n},
\end{align}
where $c_0$ is the activity of the empty sites, $c_1$ is the activity of the dimers, and $N(L,n)$ is the number of configurations of $n$ dimers and $L-2n$ empty sites on the length $L$ chain.
$N(L,n)$ is computed from the generating function by
\begin{align}
    N(L,n)=\frac{1}{(L-2n)!n!}\frac{\partial^{L-n}}{\partial c_{0}^{L-2n}\partial c_{1}^{n}}F_{L}(c_{0},c_{1})\Big|_{c_{0}=c_{1}=0.}\label{app-eq:gen-func-derivative}
\end{align}
The generating function can be computed from a recurrence relation. A chain of $L$ sites must terminate at one end with either an empty site (with activity $c_0 $) or a dimer (with activity $c_1$), hence $F_L(c_0,c_1)$ is given in terms of $F_{L-1}(c_0,c_1)$ and $F_{L-2}(c_0,c_1)$ by
\begin{align}
    F_{L}=c_{0}F_{L-1}+c_{1}F_{L-2}.
\end{align}
The initial conditions are $F_{0}=1$ and $F_{1}=c_{0}$; thus, the recurrence relation is solved by
\begin{align}
    F_{L}(c_0,c_1)=\frac{\lambda_{+}^{L+1}-\lambda_{-}^{L+1}}{\lambda_+ - \lambda_-},
\end{align}
where $\lambda_{\pm}=\frac{1}{2}(c_{0}\pm\sqrt{c_{0}^{2}+4c_{1}})$ are the roots of the characteristic equation $\lambda^{2}=c_{0}\lambda+c_{1}$.

At this point, $N(L,n)$ can be computed directly from Eq.~\eqref{app-eq:gen-func-derivative} analytically (up to a finite sum) by  expanding $F_L(c_0,c_1)$ as follows.
First, let $\Delta = \sqrt{c_0^2+4c_1}$ so that $\lambda_\pm = \frac{1}{2}(c_0 \pm \Delta)$ and $\lambda_+ - \lambda_- = \Delta$ and expand $F_L(c_0,c_1)$
\begin{align}
    F_L(c_0,c_1) = \frac{1}{2^{L+1}\Delta}\sum_{k=0}^{L+1} \binom{L+1}{k}c_0^{L+1-k}\Delta^k \left[ 1 - (-1)^k \right].
\end{align}
Note that only the odd-$k$ terms survive. Rewriting the sum to retain only odd-$k$ terms, and expanding the power of $\Delta$, we find
\begin{align}
    F_L(&c_0,c_1) =  \sum_{k=0}^{\lfloor L/2 \rfloor}\sum_{q=0}^k \binom{L+1}{2k+1} \binom{k}{q} \frac{c_0^{L-2k+2q}c_1^{k-q}}{2^{L-2k+2q}}.
\end{align}
Substituting this expression into Eq.~\eqref{app-eq:gen-func-derivative}, taking the derivative with respect to $c_1$, and evaluating at $c_1=0$,
\begin{align}
    \frac{1}{n!}\frac{\partial^n}{\partial c_1^n}c_1^{k-q}\Bigg|_{c_1=0} = \delta_{k-q,n}.
\end{align}
The sum over $q$ collapses due to the Kronecker-$\delta$ and the resulting expression as a function of $c_0$ is $(1/n!)\partial^n_{c_1}F_L|_{c_1=0} \propto c_0^{L-2n}$. 
Differentiating this expression $(L-2n)$ times with respect to $c_0$ and evaluating at $c_0 = 0$ then yields
\begin{align}
    N(L,n) = \frac{1}{2^{L-2n}} \sum_{k=n}^{\lfloor L/2\rfloor} \binom{L+1}{2k+1}\binom{k}{k-n}.
\end{align}
It is not immediately obvious that this is equivalent to Eq.~\eqref{app-eq:obc-state-norm}, but one may either trust the general combinatorial argument given in App.~\ref{app:1d-p-wave} or readily verify the equivalence numerically for any $L$ and $n\leq \lfloor L/2 \rfloor$.


\bibliography{fermion-htrs}

\end{document}